\documentclass[aps,10pt,pre,floatfix,twocolumn,superscriptaddress,showpacs]{revtex4-1}
\usepackage{color}
\usepackage{graphicx}
\usepackage{subfigure,amsmath, amsthm, amssymb}
\usepackage[colorlinks,citecolor=red]{hyperref}

\newcommand{\be}{\begin{equation}}
\newcommand{\ee}{\end{equation}}
\newcommand{\bea}{\begin{eqnarray}}
\newcommand{\eea}{\end{eqnarray}}

\newcommand{\lm}{\mathcal{L}}

\newcommand{\pbc}{\Omega_p(L)}
\newcommand{\obc}[1]{\Omega_o(#1)}

\begin{document}

\title{Phase diagram of a system of hard cubes on the cubic lattice}

\author{N. Vigneshwar}
\email{vigneshwarn@imsc.res.in}
\affiliation{The Institute of Mathematical Sciences, C.I.T. Campus,
Taramani, Chennai 600113, India}
\affiliation{Homi Bhabha National Institute, Training School Complex, Anushakti Nagar, Mumbai 400094, India}
\author{Dipanjan Mandal}
\email{mdipanjan@imsc.res.in}
\affiliation{The Institute of Mathematical Sciences, C.I.T. Campus,
Taramani, Chennai 600113, India}
\affiliation{Homi Bhabha National Institute, Training School Complex, Anushakti Nagar, Mumbai 400094, India}
\author{Kedar Damle}
\email{kedar@theory.tifr.res.in}
\affiliation{Department of Theoretical Physics, Tata Institute of Fundamental Research, Mumbai 400 005, India}
\author{Deepak Dhar}
\email{deepak@iiserpune.ac.in}
\affiliation{Indian Institute of Science Education and Research, Dr. Homi Bhabha Road, Pashan, Pune 411008, India}
\author{R.Rajesh }
\email{rrajesh@imsc.res.in}
\affiliation{The Institute of Mathematical Sciences, C.I.T. Campus,
Taramani, Chennai 600113, India}
\affiliation{Homi Bhabha National Institute, Training School Complex, Anushakti Nagar, Mumbai 400094, India}
\date{\today}

\begin{abstract}
We study the phase diagram of a system of $2\times2\times2$ hard cubes on a  three dimensional cubic lattice. Using Monte Carlo simulations, we show that the system  exhibits four different phases as the density of cubes is increased: disordered, layered, sublattice ordered, and columnar ordered. In the layered phase, the system spontaneously breaks up into  parallel slabs of size $2\times L \times L$  where only a very small fraction cubes do not lie wholly within a slab. Within each slab, the cubes are disordered; translation symmetry is thus broken along exactly one principal axis.  In the solid-like sublattice ordered phase, the hard cubes preferentially occupy one of eight sublattices of the cubic lattice,  breaking translational symmetry along all three principal directions. In the columnar phase,  the system spontaneously breaks up into weakly interacting parallel columns of size $2\times 2\times L$ where only a very small fraction cubes do not lie wholly within a column. Within each column, the system is disordered, and thus translational symmetry is broken
only along two principal directions. Using finite size scaling, we show that  the  disordered-layered phase transition is continuous, while the layered-sublattice and sublattice-columnar transitions are discontinuous. We construct a Landau theory written in terms of the layering and columnar order parameters, which is able to describe the different phases that are observed in the simulations and the order of the transitions. Additionally, our results near the disordered-layered transition are consistent with the $O(3)$ universality class perturbed by cubic anisotropy as predicted by the Landau theory.
\end{abstract}


\maketitle
\section{Introduction}

Models with only excluded volume interactions have been studied for a long time as the simplest statistical models of thermodynamic phase transitions. In these models,
the phases and phase transitions are completely determined by the shape and density of the particles, and temperature plays no role. Well-known examples include the isotropic-nematic transition in long needles~\cite{1949-o-nyas-effects,1995-oup-gp-physics}, the freezing transition in hard spheres~\cite{1957-aw-jcp-phase,1957-wj-jcp-preliminary, 2015-jcp-ik-hard}, and phase transitions in lattice models with nearest neighbour exclusion~\cite{1958-inc-d-some, 1960-b-pps-lattice, 1961-b-pps-lattice}. Experimental systems exhibiting such entropy-driven phase transitions include those between nematic, smectic and cholesteric phases in liquid crystals~\cite{1995-oup-gp-physics}, nanotube gels~\cite{2004-pre-iadzly-nematic}, and suspensions of tobacco mosaic virus~\cite{1989-prl-fmcm-isotropic}. Other examples exhibiting such transitions include adsorbed gas molecules on metallic surfaces~\cite{1985-prb-twpbe-two, 2000-ssr-psb-phase,1991-jcp-dmr-model}, as well as colloidal suspensions such as polymethyl methacrylate (PMMA) suspended in poly-12-hydroxystearic acid~\cite{1986-pm-nature-phase}. Despite a long history of study, a general understanding of the  dependence of the nature of the emergent
phases on the shapes of the particles, as well as the order of appearance of the phases with increasing density, is lacking.

The system of hard spheres in three dimensions was one of the first numerically studied systems~\cite{1957-aw-jcp-phase,1957-wj-jcp-preliminary} to show such an entropy-driven phase transition. It  undergoes a first-order transition from a fluid phase to a solid phase with face centred cubic packing~\cite{1998-gb-phystoday-simple,1997-nature-zlrmorc-crystallization}. More complicated shapes such as cubes, rhombohedra~\cite{2006-nanotechnology-pcyhh-controlled} or in general three dimensional regular polyhedra or corner-rounded polyhedra~\cite{2013-prl-ggrd-phase, 2012-jcp-mzl-freezing, 2010-pre-bst-phase}  have been studied as more realistic models for experimental self-assembling systems~\cite{2000-science-smwfm-monodisperse, 2009-pre-tj-dense, 2009-acsn-hcs-structure}, applications to drug delivery where shape of the carrier may decide its effectiveness~\cite{2007-jcr-ckm-particle}, biological material like immunoglobin~\cite{2008-jcp-lrj-polymer}, molecular logic gates~\cite{2011-acsn-smrmsahecj-manipulating,2011-prb-smsacrmehj-demonstration, 2013-acsn-gkkszsmej-contacting}, etc. Being able to predict the macroscopic material behaviour from  knowing its constituent building blocks would help to engineer the synthesis of materials with prescribed properties~\cite{2011-nmat-ae-mesophase,2012-science-deg-predictive}.

Of the non-spherical shapes, the simplest is a cube, which has the additional feature that cubes can be packed to fill all space. Theoretical studies in the continuum have focused on two cases: unoriented cubes whose faces are free to orient in any direction, and parallel hard cubes whose axes are parallel to the coordinate axes. The system of unoriented cubes was shown, using Monte Carlo and event driven molecular dynamics simulations,  to undergo a first order freezing transition from a fluid to a solid phase at a  critical packing fraction $\eta \approx 0.51$~\cite{2012-pnas-sfmd-vacancy}. Other simulations, however, found a cubatic phase that is sandwiched between the fluid and solid phases for packing fractions in the range $0.52<\eta<0.57$~\cite{2011-nmat-ae-mesophase}. It has been claimed in Ref.~\cite{2012-pnas-sfmd-vacancy} that the cubatic phase is a finite-size artifact. In the case of parallel hard cubes, early work focused on finding the equation of state using  high-density expansions~\cite{1964-jcp-h-high}, and low-density virial expansion up to the seventh virial coefficient~\cite{1956-jcp-z-virial,1962-jcp-hr-sixth}. Monte Carlo simulations show that the system of parallel hard cubes undergoes a continuous freezing transition from a disordered fluid phase to a solid phase at density $\rho\approx0.48$~\cite{1998-pre-j-melting,2001-jcp-gm-closer}. The data near the critical point are  consistent with the three-dimensional Heisenberg universality class~\cite{2001-jcp-gm-closer}. These results are consistent with theoretical predictions using density functional theory~\cite{2012-jcp-bdr-free}. Within this theory, the columnar phase is found to be not a
stable phase at high densities~\cite{2001-jcp-gm-closer,2012-jcp-bdr-free}. Thus, it would appear that parallel  hard cubes in the continuum show only one phase transition and the high density phase is crystalline.

Hard-core lattice gas models also provide interesting examples of entropy-driven phase transitions,  and like in the continuum, have rich phase diagrams. Rigorous results are known for dimer gas~\cite{1970-hl-prl-monomers}, hard triangles at full packing~\cite{1999-vn-prl-triangular},  hard hexagons~\cite{1980-b-jpa-hard}, long rods~\cite{2013-dg-cmp-nematic} and hard plates in three dimensions~\cite{2018-arxiv-dgj-plate}. For other shapes, Monte Carlo are more reliable than predictions based on approximate theories. Examples include rods~\cite{2007-gd-epl-on,2008-mlr-epl-determination,2013-krds-pre-nematic},  pentamers~\cite{2000-eb-jpa-random}, tetronimos~\cite{2009-bsg-langmuir-structure}, squares~\cite{1966-bn-prl-phase, 1967-bn-jcp-phase, 2012-rd-pre-high,1966-jcp-rc-phase,2016-ndr-epl-stability, 2017-mnr-jsm-estimating}, etc. In three dimensions, the results are much fewer and the detailed phase diagram is known only for long rods~\cite{2017-gkao-pre-isotropic, 2017-vdr-jsm-different}.  Monte Carlo simulations with local moves are often inefficient in equilibriating the system when the excluded volume is large or packing fraction is high. This puts a restriction on the kind of systems one can study. Recently, we have introduced an efficient Monte Carlo algorithm with cluster moves that has helped in overcoming these difficulties~\cite{2012-krds-aipcp-monte,2013-krds-pre-nematic,2015-rdd-prl-columnar}. We have used this to
determine the unexpectedly complex phase structure, and nature of the phase transitions
in systems like rods in two~\cite{ 2013-krds-pre-nematic} and three dimensions~\cite{2017-vdr-jsm-different}, hard rectangles~\cite{2014-kr-pre-phase, 2015-kr-epjb-phase, 2015-kr-pre-asymptotic, 2016-ndr-epl-stability, 2015-nkr-jsp-high}, discretized discs~\cite{2014-nr-pre-multiple,2016-nr-jsm-high}, Y-shaped molecules~\cite{2018-pre-mnr-phase}, as well as mixtures~\cite{2015-rdd-prl-columnar, 2015-ksr-epl-phase} of hard objects.

In this paper, we study a system of $2\times2\times2$ hard cubes on the cubic lattice using
this cluster algorithm~\cite{2012-krds-aipcp-monte,2013-krds-pre-nematic,2015-rdd-prl-columnar}. The positions of  cubes are now discrete. The differences between the phase structure of
this discrete problem and the corresponding continuum one are not well understood. Earlier Monte Carlo studies of the discrete problem~\cite{2005-p-jcp-thermodynamic} found no phase transitions for densities  upto full packing (in Ref.~\cite{2005-p-jcp-thermodynamic}, the problem of cubes correspond to $\sigma=2$). On the other hand,  for cubes with sides of length two, the approximate density functional theory predicts that there should be a transition from a disordered phase to a layered phase at low densities and from a layered phase to a columnar phase at higher densities. When the length of a side is six, the theory predicts a transition from a disordered phase to a solid, and then to two types of columnar phases~\cite{2003-jcp-lc-phase}.  Simulations of a mixture of cubes of sizes two, and four or six, show a demixing transition~\cite{1994-prl-df-evidence} in contradiction to predictions from density functional theory~\cite{2002-prl-lc-elusiveness}. However, the prediction for a pure system of cubes have not, to our knowledge, been tested in large scale simulations.

Here, we find that this system of cubes goes through four distinct phases as the density of cubes is increased: disordered, layered, sublattice ordered, and columnar ordered. In the layered phase, the system spontaneously breaks up into  parallel slabs of size $2\times L \times L$ which are preferentially occupied by cubes. Within each slab, the cubes are disordered; translation symmetry is thus broken along exactly one principal axis.  In the solid-like sublattice ordered phase, the hard cubes preferentially occupy one of eight sublattices of the cubic lattice,  breaking translational symmetry along all three principal directions. In the columnar phase,  the system spontaneously breaks up into weakly interacting parallel columns of size $2\times 2\times L$ which are preferentially occupied by cubes. Within each column, the system is disordered, and the columns break translational symmetry along
along two principal directions. By studying systems of different sizes, we argue that  the  disordered-layered phase transition is continuous, while the layered-sublattice and sublattice-columnar transitions are discontinuous. We construct a Landau theory written in terms of the layering order parameter $\mathbf{L}$
and columnar order parameter $\mathbf{C}$ which is able to describe the different phases that are observed in the simulations and the order of the transitions. Additionally, our results near the disordered-layered transition are consistent with the Landau theory prediction of scaling behaviour in the $O(3)$ universality class perturbed by cubic anisotropy.

The remainder of the paper is organized as follows. Section~\ref{sec:model} defines the model precisely and describes the grand canonical Monte Carlo scheme that is used to simulate the system. Section~\ref{sec:phasesk2} describes the different phases -- disordered, layered, sublattice ordered and columnar ordered-- that we observe in our simulations. In Sec.~\ref{sec:landau}, we  understand our simulation results in terms of a Landau theory approach. Sections~\ref{sec:transitions_1} - \ref{sec:transitions_3} characterize the different phase transitions that occur in this system. Section~\ref{sec:stability} discusses the long lived metastable states that we observe at densities close to full packing. This section also presents a perturbation expansion that allows us to argue that the high density phase will be columnar. Finally, Sec.~\ref{sec:discussion} contains a discussion of our results.

\section{\label{sec:model}Model \& Algorithm}

Consider a $L\times L\times L$ cubic lattice with periodic boundary conditions and even $L$. The lattice may be occupied by cubes of size $2\times  2 \times 2$ (i.e having side-length of $2$ lattice spacings) whose positions are in registry with the lattice sites. We associate a weight $z=e^\mu$ with each cube, where $z$ is the activity and $\mu$ is the chemical potential. The cubes interact through only excluded volume interaction, {\em i.e.} no two cubes can overlap in volume. 
For a cube, we identify the  vertex  with minimum $x$-, $y$-, and $z$-coordinates as its head. The configuration of the system can thus be fully specified by the spatial coordinates
of the heads of all the cubes in the system.
\begin{figure}
\includegraphics[width=\columnwidth]{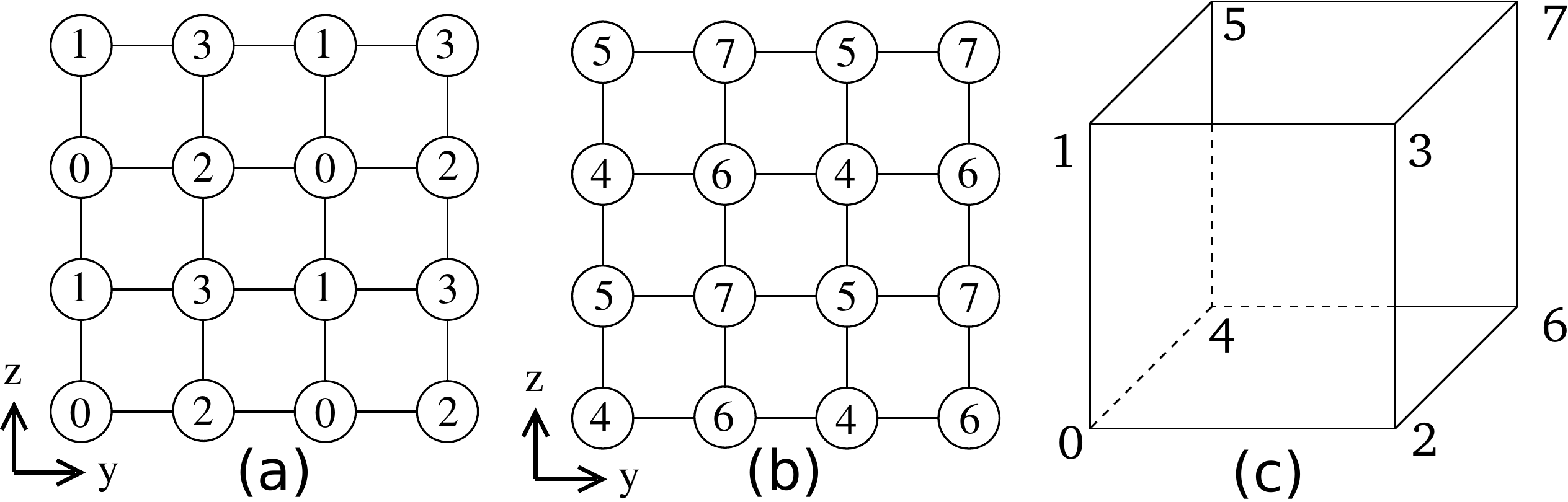}
\caption{The lattice is divided into eight sublattices $0,1,\dots,7$ depending on whether the $x$-, $y$- and $z$- coordinates are even or odd. Labeling of sublattices corresponding to $yz$-planes whose $x$-coordinate is (a) even, or  (b) odd.  (c) A $2\times2\times2$ cube with all of its vertices labelled with appropriate sublattices to show the relative positions of the planes shown in (a) and (b). }
\label{sub_label}
\end{figure}

We study the model using grand canonical Monte Carlo simulations implementing an algorithm which includes cluster moves~\cite{2012-krds-aipcp-monte,2013-krds-pre-nematic,2015-rdd-prl-columnar,2014-nr-pre-multiple} that help in equilibrating systems of hard particles with large excluded volume at densities close to full packing~\cite{2012-krds-aipcp-monte,2013-krds-pre-nematic} or at full packing~\cite{2015-rdd-prl-columnar}. Below, we briefly summarise the algorithm. 

Choose at random one of the $3L^2$ rows, where each row consists of $L$ consecutive sites in any one direction. Evaporate all the cubes whose ``heads'' lie on this row. The row now consists of empty intervals separated from each other by sites that cannot be occupied by the head of a cube due to the hard constraints arising from cubes in neighbouring rows. The empty intervals are reoccupied by new configurations of cubes with the correct equilibrium probabilities.  The calculation of these probabilities reduces to a one dimensional problem which may be solved exactly by standard transfer matrix methods (see Refs.~\cite{2013-krds-pre-nematic,2015-rdd-prl-columnar,2014-nr-pre-multiple,2017-vdr-jsm-different}  for details). This evaporation and deposition move satisfies detailed balance as the transition rates depend only on the equilibrium probabilities of the new configuration. We use a parallelised version of the algorithm described above, which exploits the fact that the rows separated from each other by a distance two can be updated independently and concurrently. We check for equilibration by taking different initial configurations of the system that correspond to different phases and confirming that the results are independent of the initial configuration. We find that the algorithm is able to equilibrate systems with density upto $\approx 0.95$ for $L \gtrsim 100$, though slightly larger densities may be attained for smaller systems.
 
\section{\label{sec:phasesk2} Different Phases}

We first define and describe the phases that we observe in our simulations. To do so, it is convenient to divide the lattice into 8 sublattices, depending on whether each $x$-, $y$-, and $z$- coordinates are odd or even. A site with coordinates $(x,y,z)$ belongs to sublattice whose binary representation is ($x$ mod $2$) ($y$ mod $2$) ($z$ mod $2$), as shown in Fig.~\ref{sub_label}. We define $\rho_i$ to be the fraction of lattice sites occupied by the cubes whose heads lie on the sublattice $i$. The total density of the system $\rho$ is then
\be
\label{eq:p}
\rho = \sum_{i=0}^7 \rho_i.
\ee

Further, let $\eta(x,y,z)$ be equal to $1$  if $(x,y,z)$ is occupied by a head of the cube, and be equal to $0$ otherwise. Consider the Fourier transform
\be
\tilde{\eta}(k_x, k_y,k_z) = \frac{8}{L^3}\,\sum_{x,y,z}\,\eta(x,y,z)\,\mathrm{e}^{i(k_x x + k_y y + k_z z)}\label{eq:fourier}.
\ee
We define the order parameter  $\mathbf L$ as
\be
\mathbf{L} = (L_x, L_y, L_z),\label{eq:L}
\ee
where $L_x = \tilde{\eta}(\pi,0,0)$, $L_y = \tilde{\eta}(0,\pi,0)$ and $L_z=\tilde{\eta}(0,0,\pi)$. A non-zero value in $L_x$ will imply that there is translational order of period two in $x$ direction. Similar interpretations hold for $L_y$ and $L_z$. The $\mathbf{L}$ vector is
thus a measure of the layering tendency of the system in each cartesian direction, and we shall refer to it as the layering vector. In a layered phase, only one cartesian component of $\mathbf{L}$ is expected to be nonzero in the thermodynamic limit. In contrast, a columnar-ordered phase is characterized by
a layering vector with two nonzero cartesian components. Finally, a solid-like sublattice-ordered phase is characterized by a layering vector with all three components
nonzero.

Note that $\mathbf{L}$ serves as a ``faithful'' order parameter for
each of these three phases: In the layered case, it correctly distinguishes
between the six symmetry-related states of the system (corresponding to
the two possible layered states for layering along each of the three cartesian directions)
by taking on the six symmetry related values $(\pm |L|, 0, 0)$, $(0,\pm |L|,0)$, and
$(0,0,\pm |L|)$.
In the columnar-ordered case, it correctly distinguishes between the twelve symmetry-related columnar states by taking on the twelve symmetry-related values  $(\pm |L|, \pm |L|, 0)$, $(0,\pm |L|,\pm |L|)$, and
$(\pm |L|,0,\pm |L|)$. Finally, the eight symmetry-related sublattice  ordered states are
correctly described by the eight symmetry-related values $(\pm |L|,\pm |L|,\pm |L|)$.

To characterize these phases, it is also useful to define two other measures of
spontanously broken symmetry: the columnar vector $\mathbf{C}$ whose
components are given by $C_x = \tilde{\eta}(0,\pi,\pi)$, $C_y = \tilde{\eta}(\pi,0,\pi)$ and $C_z=\tilde{\eta}(\pi,\pi,0)$, and the sublattice scalar $\phi =  \tilde{\eta}(\pi,\pi,\pi)$.
In contrast to $\mathbf{L}$, neither $\mathbf{C}$ nor $\phi$ fully distinguish
between the broken symmetry states in which they are nonzero
in the thermodynamic limit. This is clear since $\mathbf{C}$ is expected to
be nonzero in the columnar ordered phase and the sublattice ordered phase, but
does not correctly distinguish between the twelve symmetry-related columnar
states or the eight symmetry-related sublattice ordered states of the system. Similarly,
$\phi$ is expected to be nonzero in the sublattice ordered phase, but does not
correctly distinguish between the eight symmetry-related sublattice ordered states.

The underlying reason for this distinction between $\mathbf{L}$ on the one hand, and $\mathbf{C}$ and
$\phi$ on the other, is clarified considerably if we pass from the globally defined quantities $\mathbf{L}$, $\mathbf{C}$ and $\phi$, to the corresponding local fields $\mathbf{L}(\vec{r})$, $\mathbf{C}(\vec{r})$ and $\phi(\vec{r})$. These local fields should be thought of
as being the coarse-grained variables (coarse-grained over a linear scale of a few lattice spacings) whose sum over the entire volume gives
the corresponding global variables. Thinking in terms of these local fields,
we see that $C_x(\vec{r}) \sim L_y(\vec{r}) L_z(\vec{r})$ (and similarly for the other components). This is related to the fact that the composite variable $L_yL_z$ acts as a field
that couples linearly to $C_x$ in a Landau-type description of spontaneous symmetry breaking. Likewise, $\phi(\vec{r}) \sim L_x (\vec{r}) L_y (\vec{r})L_z(\vec{r})$. Thus, $\mathbf{C}(\vec{r})$ and $\phi(\vec{r})$ take on mean values set by {\em composite} variables formed from the components of the local layering vector, which emerges as the fundamental quantity for describing the broken symmetries of the system. It is therefore
not surprising that the corresponding global variables ${\mathbf{C}}$ and
$\phi$ do not fully distinguish between different symmetry-related states with
spontaneous columnar or sublattice order.
This also suggests that a Landau theory for all three broken symmetry phases should
involve $\mathbf{L}$ as the key variable, although we shall see below that
the symmetry-allowed couplings between $\mathbf{L}$ and $\mathbf{C}$ and $\phi$
can also play a crucial role in determining the structure of the phase diagram.

In our simulations, we measure the magnitudes of the global variables $\mathbf{L}$,
$\mathbf{C}$ and $\phi$:
\bea
q_1&=& \sqrt{L_x^2 + L_y^2 + L_z^2},\label{eq:q1}\\
q_2 &=& \sqrt{C_x^2+C_y^2+C_z^2},\label{eq:q2}\\
q_3 &=& |\phi |. \label{eq:q3}
\eea
In addition, we monitor the joint probability distribution (histogram) of $L_x$, $L_y$
and $L_z$ in order to visualize the nature of the symmetry breaking present
in various ordered states.

\begin{figure}
\includegraphics[width=\columnwidth]{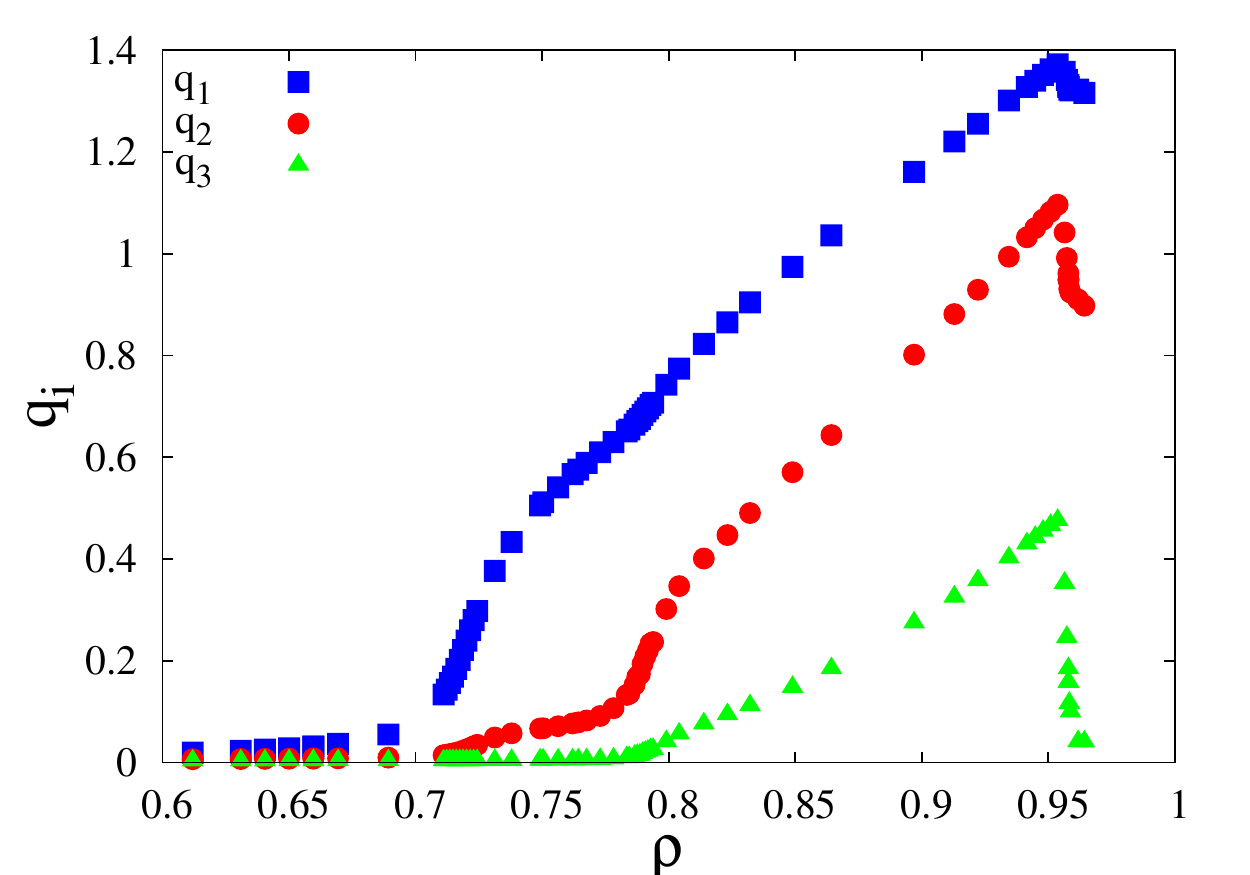}
\caption{Variation of $q_1, q_2, q_3$, as defined in Eqs.~(\ref{eq:q1})--(\ref{eq:q3}),  with density $\rho$. The data are for system size $L=70$. the discontinuities in density are not visible at this resolution.}
\label{S_vs_rho}
\end{figure}

The variation of $q_i$ with density $\rho$ is shown in Fig.~\ref{S_vs_rho}. For low densities $q_i \rightarrow 0$ in the thermodynamic limit for $i=1,2,3$ and the system is in a disordered phase. As the density is increased, $q_1$ becomes non-zero  in the thermodynamic limit
when the density crosses $\rho \approx 0.718$, signalling the onset of spontaneous layering, while $q_2$ and $q_3$  continue to be zero in the thermodynamic limit. Upon further increasing the
density, both $q_2$ and $q_3$ become
nonzero in the thermodyamic limit when the density increases beyond $\rho \approx 0.79$. This corresponds to the onset of a cystalline phase with spontaneous sublattice ordering.
Finally, when the density goes beyond $\rho \approx 0.957$, $q_3$ becomes zero, while $q_2$ and $q_1$ remain nonzero, corresponding to columnar order.
Below, we describe the behaviour of the system in each of these phases in some more detail.

{\it{Disordered Phase:}}
At low densities, the cubes are in a disordered phase in which the cubes are far apart and there is no ordering. All the mean sublattice densities are equal, {\em i.e.}, $\rho_i \approx \rho/8$, for $i=0,\dots,7$. In the disordered phase, all components of  $\mathbf{L}$ tend to zero in the limit of large system sizes [see Eq.~(\ref{eq:L}) for definition] and the system
retains all the symmetries of the underlying cubic lattice. 
\begin{figure}
\includegraphics[width=\columnwidth]{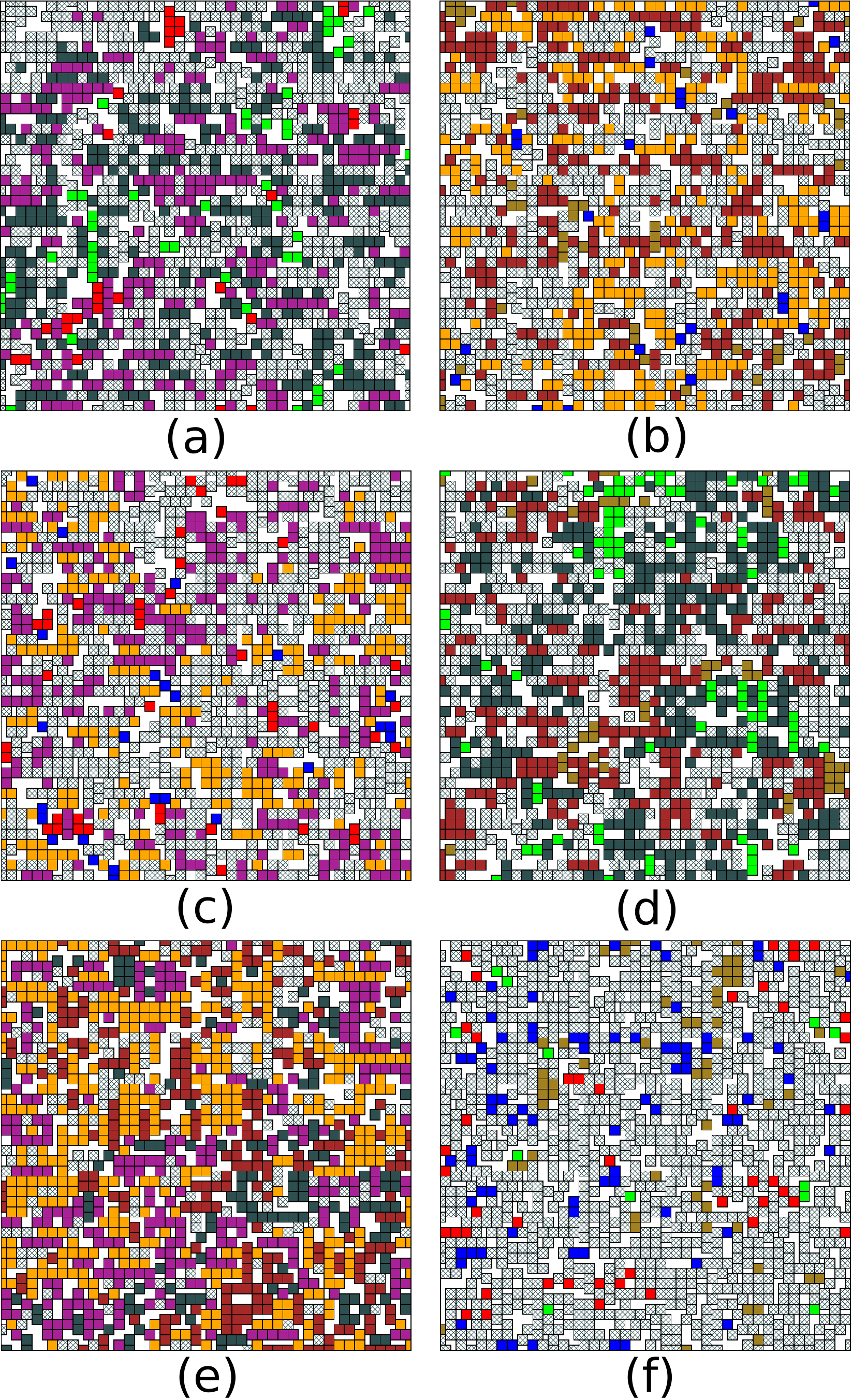}
\caption{\label{snap_layered} Snapshots of cross sections of equilibrated configurations in the layered phase, with layering vector pointing in the $z$-direction. The cross sections shown are of randomly chosen adjacent pairs of (a) even $yz$-, (b) odd $yz$-, (c) even  $xz$-, (d) odd $xz$-, (e) even $xy$- and (f) odd $xy$-planes. The eight colours represent cubes with heads on different sublattices. The projections of cubes  which protrude onto the plane from nearby planes are coloured in grey. (a)-(d) look statistically similar, while (e) is mostly coloured and (f) is mostly grey, showing a layering in the $z$-direction. The data are for system size $L=150$, chemical potential $\mu=2.4$, and  density $\rho \approx 0.762$.}
\end{figure}

{\it{Layered Phase:}}
In the layered phase, translational symmetry is broken in only one direction. The cubes preferentially occupy either odd or even planes normal
to this direction. This may be seen by examining snapshots of randomly chosen pairs of even and odd planes in the three directions as shown in Fig.~\ref{snap_layered}, where the eight different colours represent cubes whose heads on a particular sublattice. Grey colour represents sites that are occupied by cubes whose heads are on neighbouring planes. In Fig.~\ref{snap_layered}(a)-(d), showing the snapshots of randomly chosen even and odd $yz$ and $xz$ planes, there are approximately equal number of coloured cubes and grey cubes, showing both odd and even $yz$- and $xz$-planes are equally occupied. On the other hand, it can be seen that Fig.~\ref{snap_layered}(e), showing snapshot of a randomly chosen even $xy$ plane, has much larger number of coloured squares than grey squares, while Fig.~\ref{snap_layered}(f), showing snapshot of a randomly chosen odd $xy$ plane, is mostly grey, showing that in this configuration, the heads of cubes preferentially occupy even $xy$-planes.

\begin{figure}
\includegraphics[width=\columnwidth]{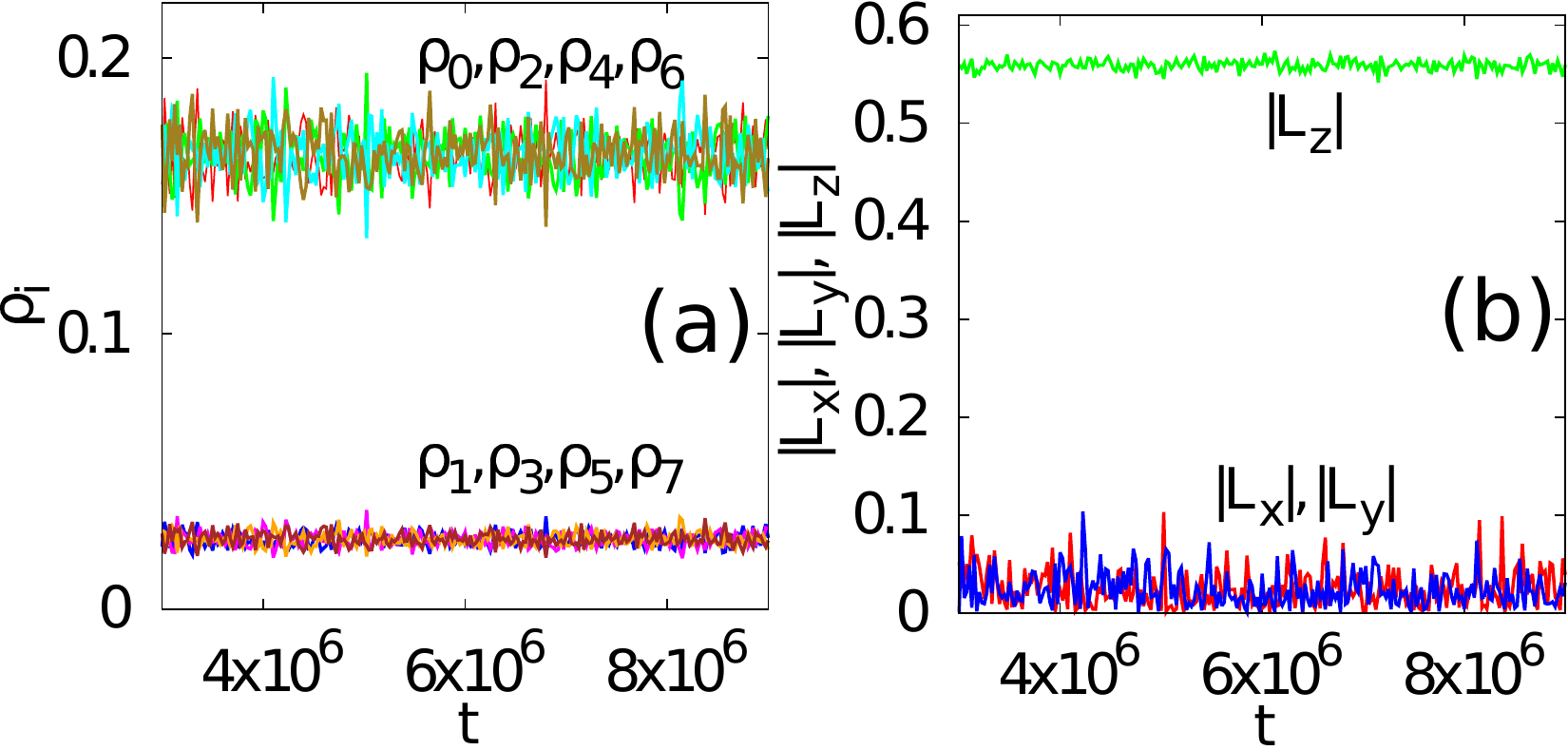}
\caption{\label{smec}Temporal evolution of (a) eight sublattice densities $\rho_i,i=0,\dots,7$ and (b) $|L_x|, |L_y|, |L_z|$ when the system is in a layered phase  (layering in the $z$-direction). The data are for  $\mu=2.4$, $\rho\approx0.762$, and system size $L=150$.} 
\end{figure}

The breaking of translational invariance is also quantitatively reflected from the time evolution  of the eight sublattice densities and $L_x, L_y, L_z$, as shown in Fig.~\ref{smec}(a) and (b) respectively. From Fig.~\ref{smec}(a), we see that four sublattices are preferentially occupied. From Fig.~\ref{smec}(b), we also see that one of the components of $\mathbf L$ is larger than the other two, i.e., $|L_z| \gg |L_x|\approx |L_y|$, confirming that the system is layered in the $z$-direction. More precisely, one component of $\mathbf{L}$ remains
nonzero in the thermodynamic limit, and the other two are zero.

\begin{figure}
\includegraphics[width=\columnwidth]{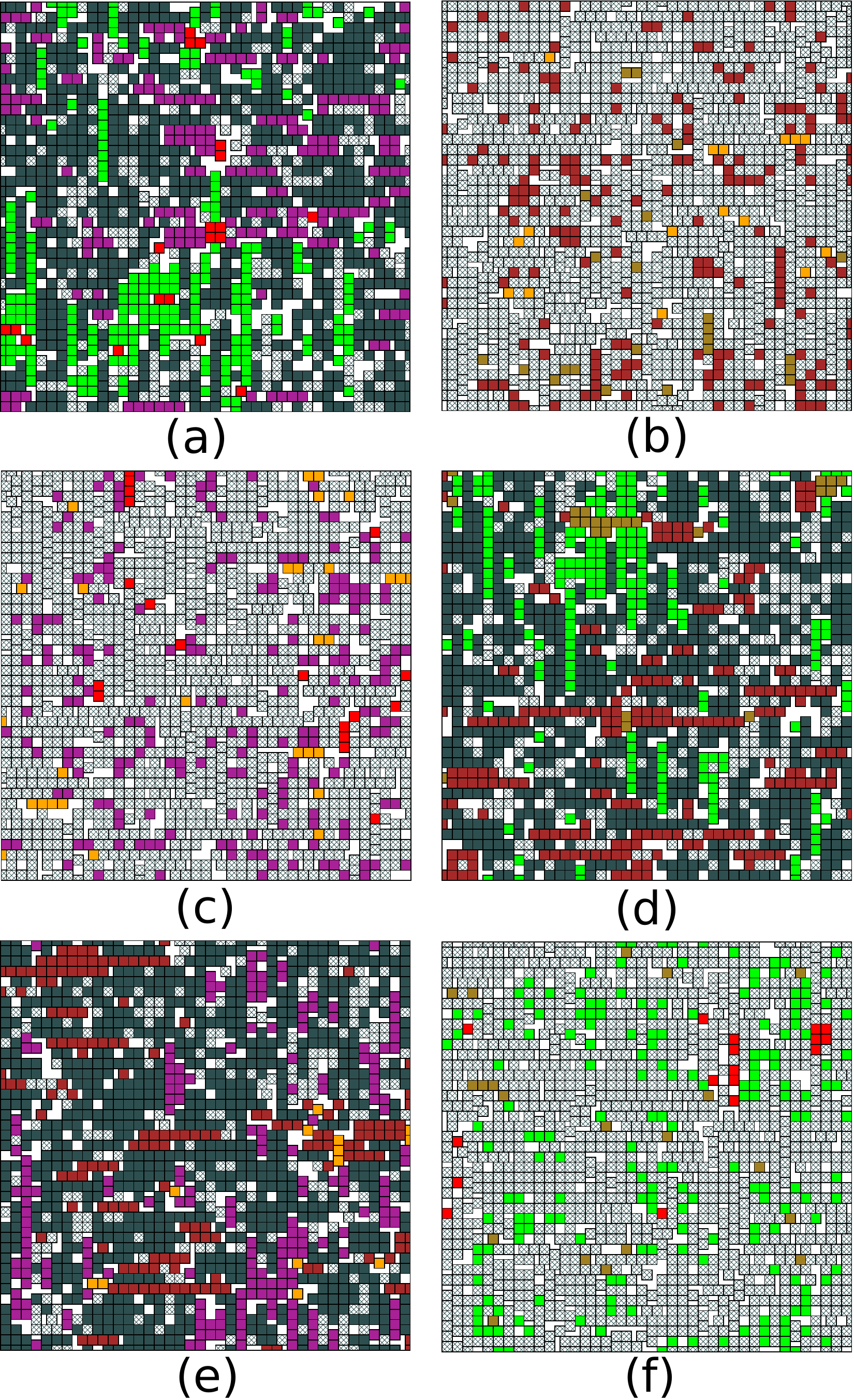}
\caption{\label{snap_sublattice} Snapshot of cross sections of equilibrated sublattice phase, where the cross sections are of randomly chosen adjacent pairs of (a) even $yz$-, (b) odd $yz$-, (c) even  $xz$-, (d) odd $xz$-, (e) even $xy$- and (f) odd $xy$-plane. The eight colours represent cubes with heads on different sublattices. The projections of cubes  which protrude onto the plane from nearby planes are coloured in grey. (a), (d) and (e) are mostly coloured by deep-green, while (b), (c) and (f) are mostly grey, showing the preferential occupancy of cubes in sublattice $2$. The data are for system size $L=150$ with chemical potential $\mu=3.5$, and  density $\rho \approx 0.864$.}
\end{figure}

{\it{Sublattice Phase:}}
In the sublattice phase,  translational symmetry is broken in all three principal directions of the cubic lattice. In this phase, the cubes preferentially occupy one of the eight sublattices. This may be seen by examining the snapshots of randomly chosen pairs of even and odd planes in the three directions as shown in Fig.~\ref{snap_sublattice}. It may be seen that in each of the directions, one of the planes has a larger number of cubes, compared to the grey squares. We see that in this case the cubes preferentially occupies simultaneously even $yz$, odd $xz$ and even $xy$-planes, which implies most of the cubes occupy sublattice 2.

\begin{figure}
\includegraphics[width=\columnwidth]{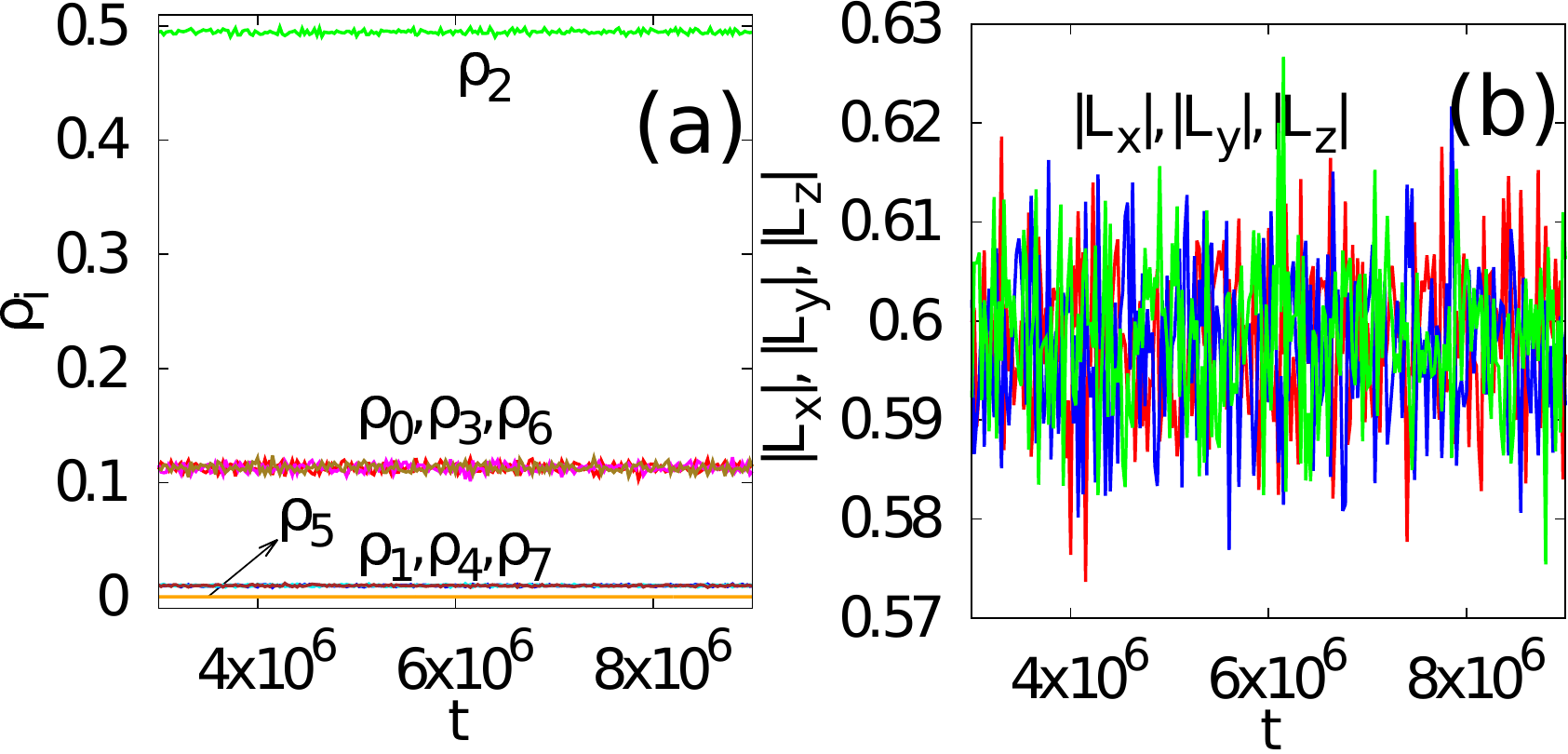}
\caption{\label{sublat} Temporal evolution of (a) eight sublattice densities $\rho_i,i=0,\dots,7$ and (b) $|L_x|, |L_y|, |L_z|$  when the system is in a sublattice phase. The data are for $\mu=3.5$, $\rho\approx0.864$, and system size $L=150$.} 
\end{figure}

The breaking of translational symmetry is reflected in the time evolution of the eight sublattice densities and $|L_x|, |L_y|, |L_z|$  as shown in Fig.~\ref{sublat}(a) and (b) respectively. In Fig.~\ref{sublat}(a), sublattice $2$ is preferentially occupied over the seven. From Fig.~\ref{sublat}(b), we also see that $|L_x|, |L_y|, |L_z|$  are non-zero and equal, i.e., $|L_x| \approx |L_y|\approx |L_z|\gg0$, confirming that the system has sublattice order.

\begin{figure}
\includegraphics[width=\columnwidth]{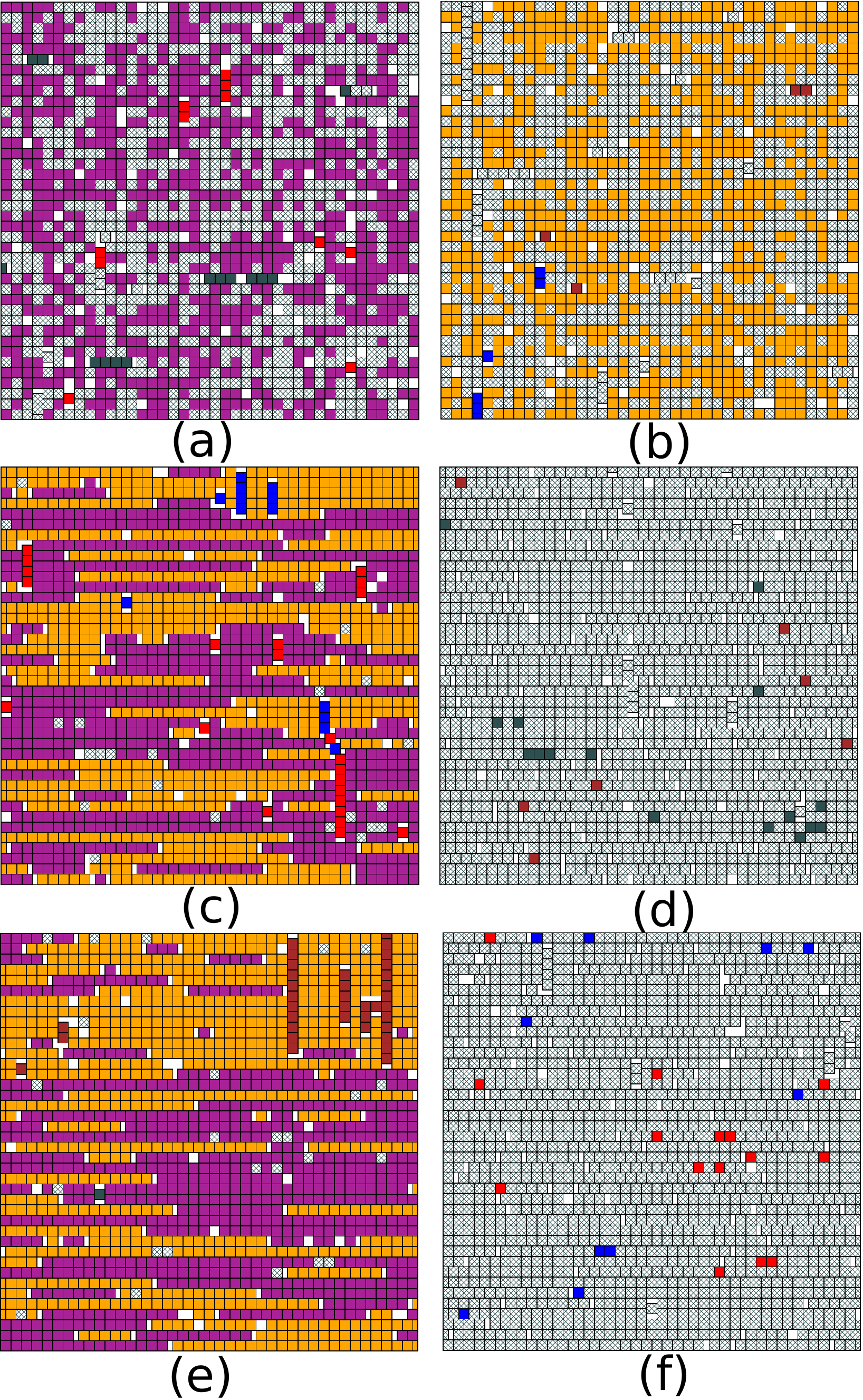}
\caption{\label{snap_columnar}Snapshot of cross sections of an equilibrated columnar phase, where the columns are aligned in the $x$-direction and $y$- and $z$- coordinates are both mostly even.  The cross sections are of randomly chosen adjacent pairs of (a) even $yz$-, (b) odd $yz$-, (c) even  $xz$-, 
(d) odd $xz$-, (e) even $xy$-, and (f) odd $xy$-plane. The eight colours represent cubes with heads on different sublattices. The projections of cubes  which protrude onto the plane from neighbouring planes are coloured in gray. Since (d) and (f) are mostly gray, the heads of most of the cubes have even $y$- and $z$- coodinates. Since (a) and (b) have roughly equal number of coloured squares, the heads of the cubes could have, with equal probability, either even or odd $x$-coorindates.  The data are for system size $L=150$, chemical potential $\mu=5.5$, and  density $\rho \approx 0.958$.}
\end{figure}

{\it{Columnar Phase:}}
The system is in a columnar phase at large densities. In the columnar phase, the system breaks translational symmetry along two directions and the heads of the cubes preferentially occupy two sublattices. This may be seen by examining the  snapshots of the planes in the three directions, as shown in Fig.~\ref{snap_columnar}. From Figs.~\ref{snap_columnar}(d) and (f), corresponding to snapshots of odd $xz$-plane and odd $xy$-plane respectively, it may be seen that these planes contain very few heads of cubes. Thus, most cubes have heads with even $y$-coordinate and even $z$-coordinate. If now the $x$-coordinate has no definite parity, then the phase will be columnar, else it will be a sublattice phase. From the snapshots of even and odd $yz$-planes, shown in  Figs.~\ref{snap_columnar}(a) and (b), it can be seen that both planes have roughly equal number of heads of cubes, showing that the $x$-coordinate has no definite parity. This feature may also be observed from the snapshots shown in Figs.~\ref{snap_columnar}(c) and (e) of even $xz$- and even $xy$- planes, where two colours are seen in each snapshot corresponding to even $x$ and odd $x$.

\begin{figure}
\includegraphics[width=\columnwidth]{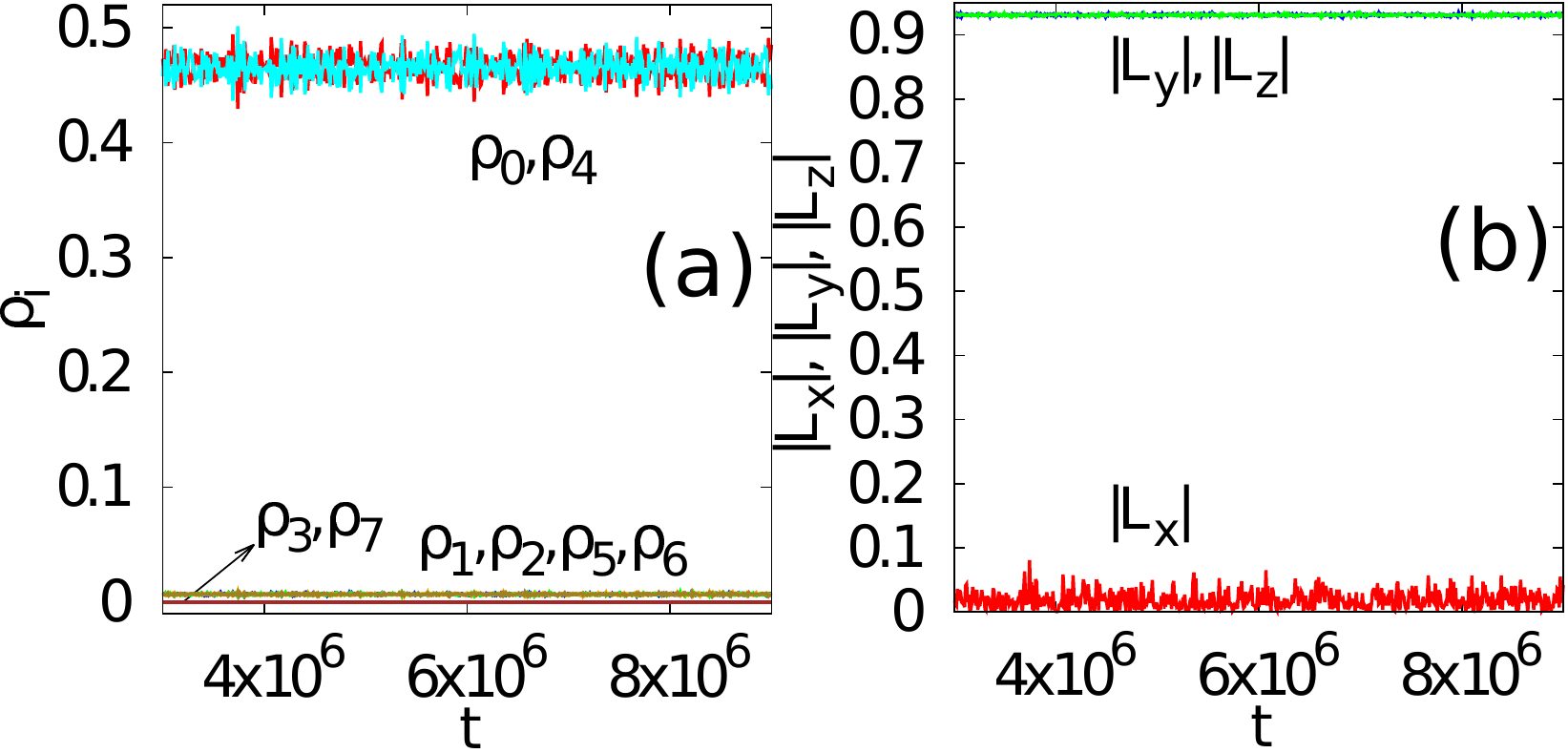}
\caption{\label{columnar}Temporal evolution of (a) the eight sublattice densities $\rho_i,i=0,\dots,7$ and (b) $|L_x|, |L_y|, |L_z|$  when the system is in a columnar phase. The data are for  $\mu=5.5$, $\rho\approx0.958$, and system size $L=150$.} 
\end{figure}

From Fig.~\ref{columnar}(a), we see that the sublattice $0$ and $4$ are preferentially occupied over the six sublattices corresponding to the heads of most of the cubes having odd $y$ and $z$-coordinates. From Fig.~\ref{columnar}(b), we see that two order parameters are large compared to one, i.e., $|L_y| \approx |L_z|\gg |L_x|$, this implies that translation symmetry is broken in both the $y$- and $z$-directions.  The columnar phase can be visualised as  a set of tubes extending along the $x$-direction, in which the cubes can slide along.

\section{\label{sec:landau} Landau theory for $2\times2\times2$ cubes}

In this section, we formulate a Landau-type theoretical description of  the phases found in Sec.~\ref{sec:phasesk2}. As noted earlier, spontaneous layering, sublattice ordering
and columnar ordering are faithfully described by the layering vector $\mathbf{L}$, defined in Eq.~(\ref{eq:L}), with the columnar vector $\mathbf{C}$ and the sublattice scalar $\phi$
more naturally thought of as composite objects constructed from the local
layering order parameter field. It is therefore natural to try to construct the Landau theory in terms of the order parameter vector $\mathbf{L}$. Here, we demonstrate
that while such a Landau theory correctly captures the low density
disordered phase, the layered phase, and the sublattice phase, as well as phase transitions between them, it  does not allow for the possibility of a columnar phase. To account
for the columnar phase, we include the symmetry-allowed couplings
to the columnar vector $\mathbf{C}$ and write down a coupled
theory for $\mathbf{L}$ and $\mathbf{C}$. This augmented Landau theory correctly
predicts the existence of a columnar phase, as well as the nature of the transition
to the columnar phase.

We start by constructing the functional only in terms of $\mathbf{L}$. The symmetries of the Landau functional $\mathcal{F}(\{L_\alpha\})$ are that it is invariant under  $\{L_\alpha \leftrightarrow -L_\alpha\}$ for $\alpha = (x,y,z)$ and cyclical permutations of the indices $(x,y,z)$. With these constraints, the most general functional is 
\begin{equation}
  \mathcal{F} = a_L|\mathbf{L}|^2 + b_L|\mathbf{L}|^4 + 2\lambda_L(L_x^2L_y^2 + L_z^2L_y^2+ L_x^2L_z^2),\label{eq:landau}
\end{equation}
where we have truncated the expansion upto fourth order.  To make sure that $\mathcal{F}$ goes to $+\infty$ when $|\mathbf{L}|\rightarrow\infty$, we require that $b_L>0$ and $\lambda_L>-3b_L/2$. The Landau theory in Eq.~(\ref{eq:landau}) is that of  $O(3)$ model with a cubic anisotropy.

The equilibrium phase is obtained from the global minimum of $\mathcal F$, and is obtained from the solutions of $\nabla_L \mathcal F=0$.  In component form, these equations are
\begin{eqnarray}
  2a_LL_x + 4b_LL_x^3 + 4L_x(b_L+\lambda_L)(L_y^2 + L_z^2) &=& 0,\label{eq:sx}\\
  2a_LL_y + 4b_LL_y^3 + 4L_y(b_L+\lambda_L)(L_x^2 + L_z^2) &=& 0,\label{eq:sy}\\
  2a_LL_z + 4b_LL_z^3 + 4L_z(b_L+\lambda_L)(L_y^2 + L_x^2) &=& 0.\label{eq:sz}
\end{eqnarray}
The solutions to Eqs.~(\ref{eq:sx})--(\ref{eq:sz}) may be found in closed form. We find that the solutions are of the form  $(0,0,0)$, $(l,0,0)$, $(s,s,s)$ and $(c,c,0)$ or its cyclic permutations. Substituting into Eqs.~(\ref{eq:sx})--(\ref{eq:sz}), the equations satisfied by $l, s, c $ are
\bea
2b_Ll^2+a_L&=&0,\label{eq:l}\\
(6b_L+4\lambda_L)s^2+a_L&=&0,\label{eq:s}\\
(4b_L+2\lambda_L)c^2+a_L&=&0.\label{eq:c}
\eea
The stability of the phases is determined by examining the Hessian $\mathcal{H}(\mathbf{L}_0)$ defined as 
\begin{equation}
  \mathcal{H}(\mathbf{L_0})_{\alpha\beta} = \left.\frac{\partial^2\mathcal{F}}{\partial L_\alpha\partial L_\beta}\right|_{\mathbf{L}=\mathbf{L}_0},\label{eq:hessian}
\end{equation}
where $\alpha$ and $\beta$ run over the indices $(x,y,z)$. For $\mathbf{L}_0$ to be local minimum or locally stable, we require that the three eigenvalues of the Hessian, calculated at $\mathbf{L}_0$,  are all positive.

For the disordered phase $(0,0,0)$, the Hessian is
\be
 \mathcal{H}(0,0,0)=
  \begin{bmatrix}
    2a_L & 0 & 0 \\
   0 & 2a_L & 0\\
  0 & 0 & 2a_L
  \end{bmatrix},
  \label{eq:dis}
\ee
whose three eigenvalues are all equal to $2a_L$. For the eigenvalues to be positive, we require that $a_L >0$.

For the layered solution $(l,0,0)$, the Hessian is
\be
 \mathcal{H}(l,0,0)=
  \begin{bmatrix}
  -4a_L & 0 & 0 \\
   & &\\
   0 & -\frac{2a_L\lambda_L}{b_L}& 0\\
    & &\\
  0 & 0 & -\frac{2a_L\lambda_L}{b_L}
  \end{bmatrix},
  \label{eq:lay}
\ee
whose eigenvalues are the diagonal entries in Eq.~(\ref{eq:lay}). For the eigenvalues to be positive, we require that $a_L<0$ and $\lambda_L>0$. 

For the sublattice solution  $(s,s,s)$, the Hessian is
\be
 \mathcal{H}(s,s,s)\!=
  \begin{bmatrix}
  -\frac{4a_Lb_L}{3b_L+2\lambda_L} & -\frac{4a_L(b_L+\lambda_L)}{3b_L+2\lambda_L} & -\frac{4a_L(b_L+\lambda_L)}{3b_L+2\lambda_L} \\
  & &\\
  -\frac{4a_L(b_L+\lambda_L)}{3b_L+2\lambda_L} &  -\frac{4a_Lb_L}{3b_L+2\lambda_L} & -\frac{4a_L(b_L+\lambda_L)}{3b_L+2\lambda_L}\\
   & &\\
  -\frac{4a_L(b_L+\lambda_L)}{3b_L+2\lambda_L}& -\frac{4a_L(b_L+\lambda_L)}{3b_L+2\lambda_L} &  -\frac{4a_Lb_L}{3b_L+2\lambda_L} 
  \end{bmatrix},
  \label{eq:sub}
\ee
whose eigenvalues are $-4a_L$, $4a_L\lambda_L/(3b_L+2\lambda_L)$, and $4a_L\lambda_L/(3b_L+2\lambda_L)$.  For the eigenvalues to be positive, we require that $a_L<0$ and $\lambda_L<0$. 
 
For columnar solution $(c,c,0)$, the Hessian is
\be
 \mathcal{H}(c,c,0)=
  \begin{bmatrix}
  -\frac{4a_Lb_L}{2b_L+\lambda_L} & -\frac{4a_L(b_L+\lambda_L)}{2b_L+\lambda_L} & 0 \\
   & &\\
   -\frac{4a_L(b_L+\lambda_L)}{2b_L+\lambda_L} & -\frac{4a_Lb_L}{2b_L+\lambda_L}& 0\\
    & &\\
  0 & 0 & -\frac{2a_L\lambda_L}{2b_L+\lambda_L}
  \end{bmatrix},
  \label{eq:col}
\ee
whose  eigenvalues are  $-4a_L$, $4a_L\lambda_L/(2b_L+\lambda_L)$, and $-2a_L\lambda_L/(2b_L+\lambda_L)$. The ratio of the second and third eigenvalues is $-2$. This implies that the three eigenvalues cannot be made simultaneously positive. Thus, the columnar solution is not a stable solution.

From the above analysis, we find that there exists a unique stable solution for each choice of $a_L$ and $\lambda_L$. For $a_L>0$ and any $\lambda_L$,  the only stable phase is the disordered phase where $\mathbf L=0$. For $a_L<0$ and $\lambda_L>0$, we find that stable solution is a layered phase, where $\mathbf L$ is a one-component vector of the form $(l,0,0)$. For the case where $a_L<0$ and $\lambda_L<0$, the stable solution is a sublattice phase, where $\mathbf L$ is  vector of the form $(s,s,s)$.

\begin{figure}
\includegraphics[width=0.7\columnwidth]{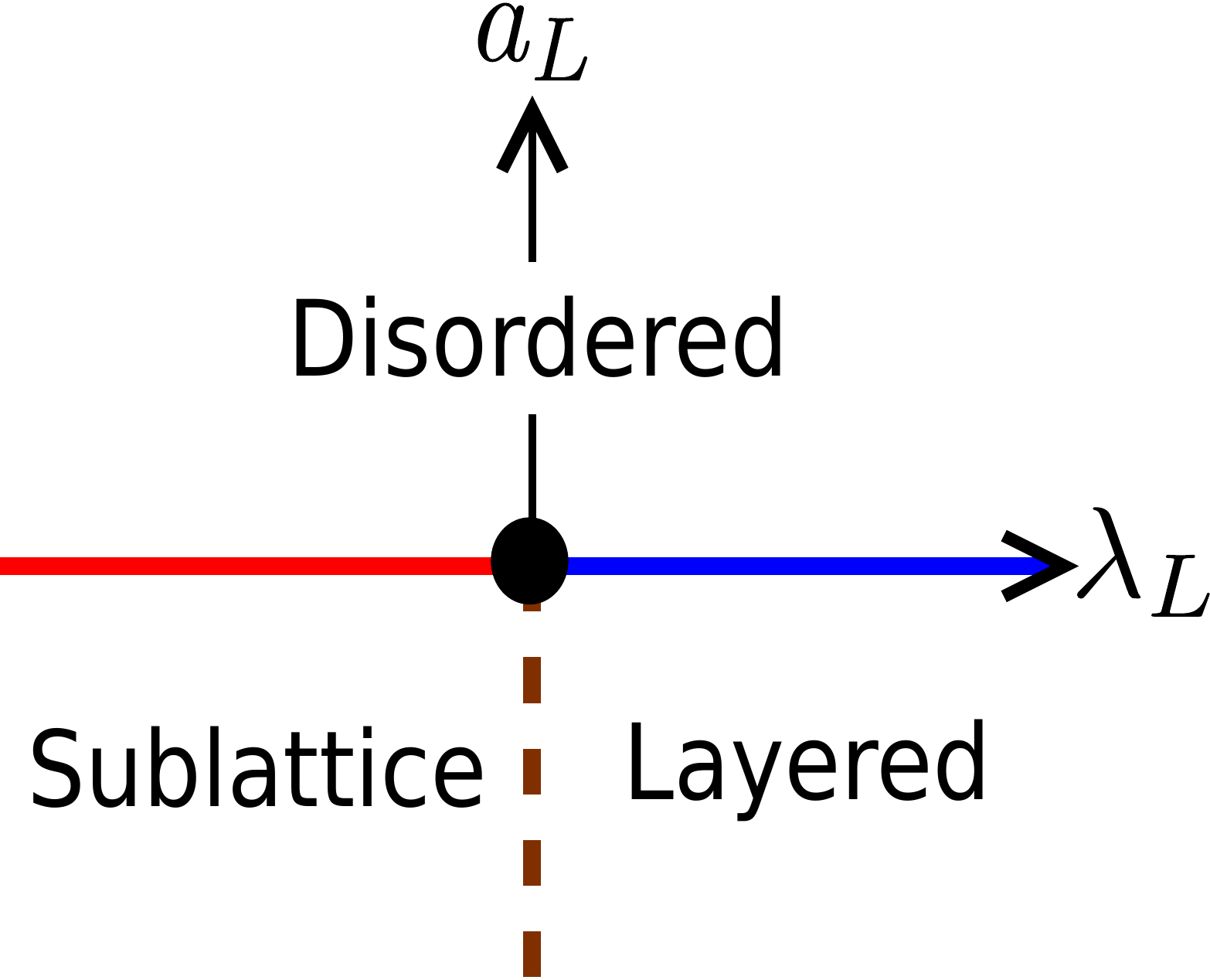}
\caption{\label{phase_diag}Phase diagram in the $\lambda_L$-$a_L$ plane for the Landau theory of Eq.~(\ref{eq:landau}). The thick red and blue lines represent lines of continuous transition, whereas the dotted brown line is a first order transition line. The three phases meet at the multicritical point $(0,0)$.} 
\end{figure}

These observations are summarized in the phase diagram  shown in Fig.~\ref{phase_diag}. The disordered-layered transition and disordered-sublattice transitions are both continuous and, within Landau theory, belong to the universality class of the $O(3)$ model with cubic anisotropy. On the other hand, the sublattice-layered transition is discontinuous, as the orientation of the $\mathbf L$ vector changes abruptly from along one of the axes to $(1,1,1)$ or an equivalent direction. 

The simplest Landau-type theory described in Eq.~(\ref{eq:landau}) predicts disordered,  layered and sublattice phases, but disallows a columnar phase. To construct a minimal theory that predicts all the phases that are seen in the simulations, we extend the functional in Eq.~(\ref{eq:landau}) to explicitly depend on  the columnar vector $\mathbf{C}(\vec{r})$ defined earlier.
Here, $\mathbf{C}(\vec{r})$ should be thought of as an independent coarse-grained vector field since,  $\langle C_x(\vec{r}) \rangle = \left \langle  L_y(\vec{r}) L_z(\vec{r}) \right \rangle $,  cannot
be fully determined in terms of $\langle L_y(\vec{r}) \rangle$ and $\langle L_z(\vec{r}) \rangle$.
The  Landau functional  $\mathcal{F}(\{L_\alpha, C_\alpha\})$ should be invariant under  $L_\alpha \leftrightarrow -L_\alpha$, and $C_\alpha \leftrightarrow -C_\alpha$ for $\alpha = (x,y,z)$ and under cyclical permutations of the indices $(x,y,z)$. The augmented functional, truncated upto fourth order is:
\begin{eqnarray}
  \mathcal{F} &=& a_L|\mathbf{L}|^2 + b_L|\mathbf{L}|^4 + 2\lambda_L(L_x^2L_y^2 + L_z^2L_y^2+ L_x^2L_z^2)\nonumber \\
  &&+a_c|\mathbf{C}|^2 + b_c|\mathbf{C}|^4+ 2\lambda_c(C_x^2C_y^2 + C_z^2C_y^2+ C_x^2C_z^2) \nonumber\\
  && - \mu(C_xL_yL_z+C_yL_xL_z+C_zL_xL_y), \label{eq:landau_coupled}
\end{eqnarray}
where $\mu$ couples the $\mathbf L$ and $\mathbf C$ vectors. We restrict ourselves to $\mu >0$ since this correctly describes the situation in a columnar ordered configuration
of our system of cubes (with our definitions of these vectors, it is easy to
see that $C_x$ has the same sign as the product $L_y L_z$ in a columnar ordered
state with columnar vector pointing in the $x$ direction, and similarly for columnar vectors
pointing in the other cartesian directions). As we demonstrate below, this augmented Landau theory now accounts for the presence of a stable columnar phase in addition
to the other stable phases already obtained by thinking entirely in terms of $\mathbf{L}$

The extended functional in Eq.~(\ref{eq:landau_coupled}) has seven independent parameters and six variables. Deriving analytic equations of the phase boundary is not possible as that would require us to simultaneously solve six coupled equations. Rather, we focus on showing that there are parameter regimes for which the columnar phase, as well as the other phases exist and are stable.  This  is achieved by assigning numerical values to the  parameters values and solving the coupled equations for equilibrium numerically. The stability is checked using the 
Hessian $\mathcal{H}(\mathbf{L}, \mathbf{C})$:
\begin{equation}
  \mathcal{H}(\mathbf{L}_0, \mathbf{C}_0)_{\alpha\beta} = \left.\frac{\partial^2\mathcal{F}}{\partial \phi_\alpha\partial \phi_\beta}\right|_{\mathbf{L}=\mathbf{L}_0, \mathbf{C}=\mathbf{C}_0},\label{eq:hessian_new}
\end{equation}
where $\phi$ runs over the components of the vectors $\mathbf{L}$ and $\mathbf{C}$, is now a $6\times6$ matrix.

For analysing the  functional in Eq.~(\ref{eq:landau_coupled}) to determine its global minima, we consider the four different cases discussed below. For each of these cases, we set $\mu=2$ and fix  the parameters $b_L$ and $b_c$ to be large and positive ($b_L=b_c=8$).

Case 1. $a_L>0$, $a_c>0$: In this case, in the absence of the coupling ($\mu=0$), both $\mathbf{L}=0$, and $\mathbf{C}=0$. For small positive $\mu$, we expect the system to be still in the disordered phase. We confirm this by setting  $a_L=a_c=1.2$, and treating $\lambda_L$ and $\lambda_c$ as free parameters. For  this case, whatever be the values and sign of $\lambda_L$ and $\lambda_c$, we find that the disordered phase is the only stable phase.

Case 2. $a_L>0$, $a_c<0$: In this case, we expect that $\mathbf{L}=0$, and $\mathbf{C}\neq 0$. Such solutions are unphysical, and we expect that the mapping from the microscopic variables of the model to the parameters of the Landau theory is such that,  this regime is never reached.

\begin{figure}
\includegraphics[width=0.7\columnwidth]{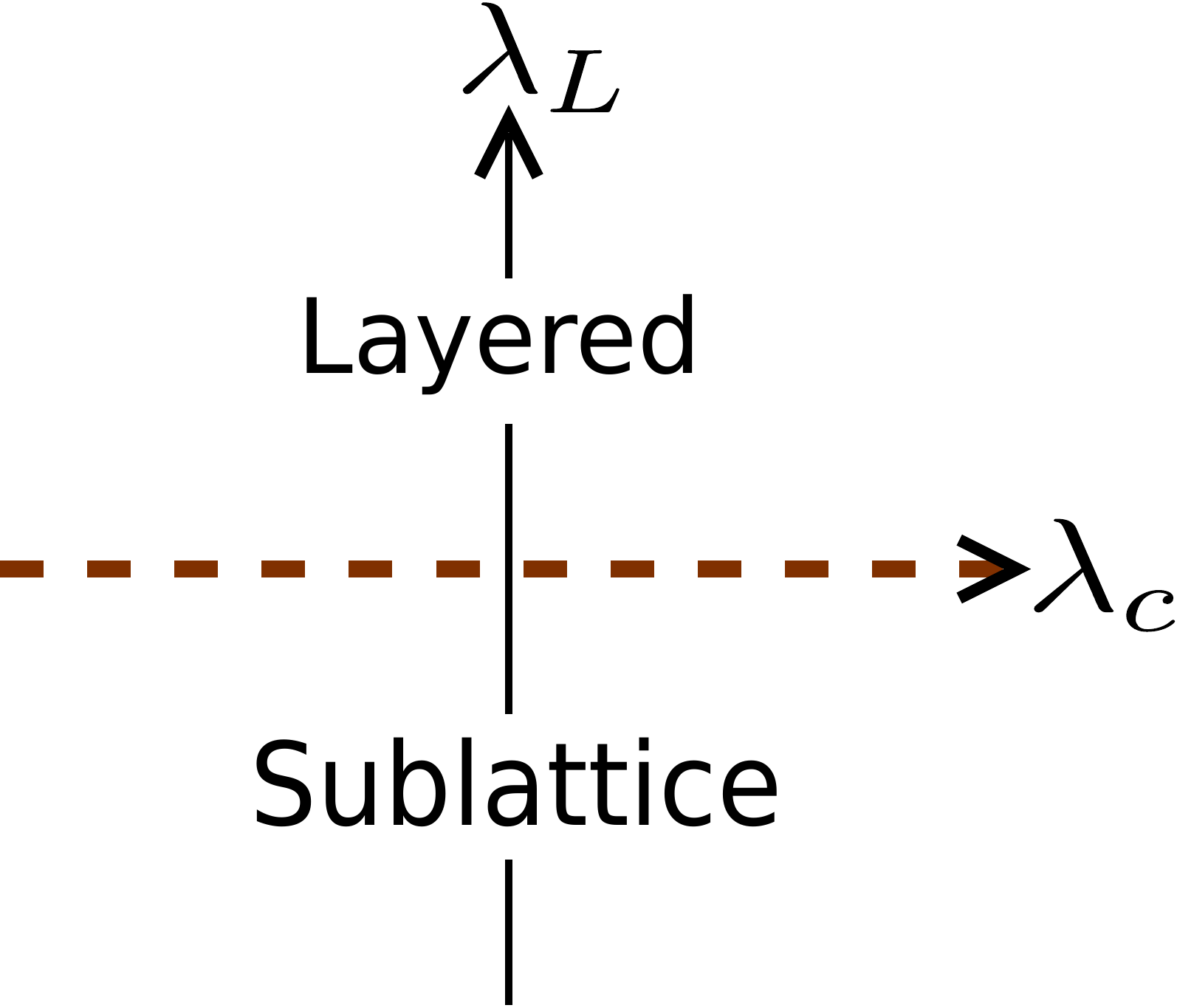}
\caption{\label{phase_diag_2}The schematic phase diagram in the $\lambda_L$-$\lambda_c$ plane for the Landau free energy functional in Eq.~(\ref{eq:landau_coupled}) for the case $a_L<0,~a_c>0$. The other parameters are $b_L=b_c=8,~\mu=2$. The line $\lambda_L=0$ is a first order line, separating the layered phase and the sublattice phase. }
\end{figure}

Case 3. $a_L<0$, $a_c>0$: In this case, in the absence of coupling ($\mu=0$), the $\mathbf{L}$ shows both layered and sublattice phases depending on the sign of $\lambda_L$. When $\mu \neq 0$, for these phases to be valid, $\mathbf{C}$ should be zero in the layered phase and have three non-zero components in the sublattice phase. We confirm that this is indeed the case by determining numerically the phase diagram for $a_L=-1.2$, $a_c=1.2$. We find that the system is layered when $\lambda_L>0$, and has sublattice order when $\lambda_L<0$, irrespective of the sign of $\lambda_c$. The schematic 
$\lambda_L$-$\lambda_c$ phase diagram for this case, obtained by minimising the free energy at different sample phase points, is summarised in  Fig.~\ref{phase_diag_2}. 

\begin{figure}
\includegraphics[width=0.7\columnwidth]{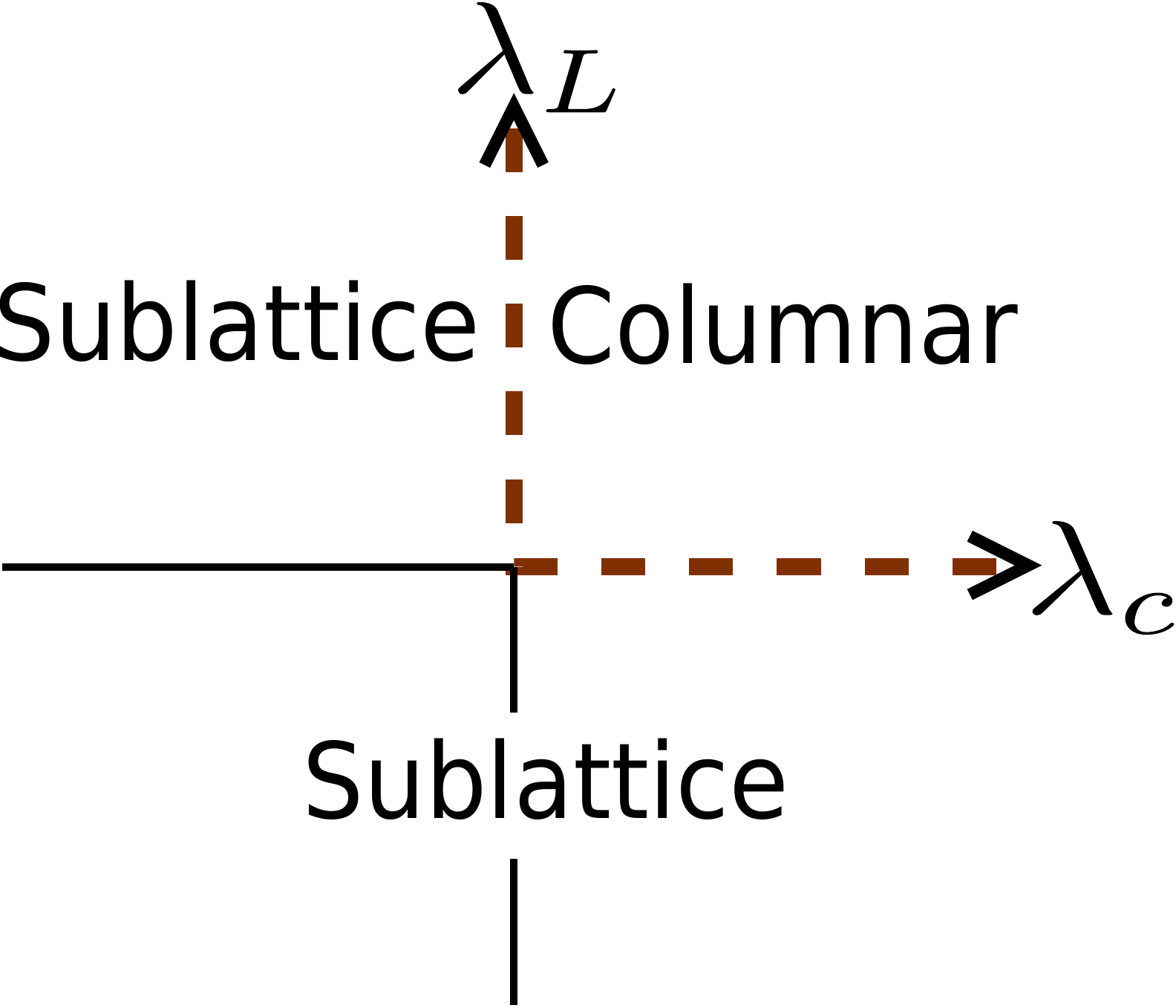}
\caption{\label{phase_diag_3}The schematic phase diagram in the $\lambda_L$-$\lambda_c$ plane for the Landau free enrgy functional in Eq.~(\ref{eq:landau_coupled}) for the case $a_L<0,~a_c<0$. The other parameters are $b_L=b_c=8,~\mu=2$. The dotted lines which are lines of first order transition, separate the columnar phase and the sublattice phase. }
\end{figure}

Case 4. $a_L<0$, $a_c<0$: In this case, we expect that when $\lambda_c>0$, then $\mathbf{C}$ could exist in a layered phase, making it possible for the system to be in a columnar phase. We determine the phase diagram for $a_L=a_c=-1.2$.  We  find that for $\lambda_c>0$ and $\lambda_L>0$, there is a regime where the system is in a columnar phase. For large $\lambda_c$ and $\lambda_L$, there are spurious unphysical solutions. For the other cases, the system is in a sublattice phase. The schematic $\lambda_L$-$\lambda_c$ phase diagram for this case,  obtained by minimising the free energy at different sample phase points, is summarised in Fig.~\ref{phase_diag_3}.

\section{\label{sec:transitions_1}Disordered-Layered Transition}

In this section, we describe the phase transition from disordered phase to layered phase. To do so, we first define susceptibility $\chi_i$, and Binder cumulant $U_i$ associated  with the order parameter $q_i$ [see Eq.~(\ref{eq:q1})--(\ref{eq:q3})  for definition] as
\bea
\label{order_parameter}
\chi_i&=&L^3(\langle q_i^2 \rangle-\langle q_i \rangle^2),\\
U_i &=& 1 - \frac{c_i \langle q_i^4 \rangle}{ \langle q_i^2 \rangle^2},
\label{cumulant}
\eea
where $c_1=c_2=9/15$ and $c_3=1/3$. The values of $c_i$ are chosen so that the Binder cumulant
is zero in the disordered phase.
We also define the deviation from the critical point as
\be
\epsilon=\mu-\mu_c,
\ee
where $\mu_c$ is the critical value of the chemical potential.

We find that the disordered-layered transition is continuous. A suitable order parameter to study this transition is $q_1$ which
is zero in the disordered phase and non-zero in layered phase. The critical behaviour may be obtained by studying the
non-analytic behaviour of the different physical quantities, which, near the transition,  is captured by the finite
size scaling behaviour:
\bea
q_1(\epsilon,L) &\simeq& L^{-\beta/\nu} f_q(\epsilon L^{1/\nu})\label{eq:fq},\\ 
\chi_1(\epsilon,L) &\simeq& L^{\gamma/\nu} f_\chi(\epsilon L^{1/\nu})\label{eq:fx},\\
U_1(\epsilon,L) &\simeq&f_U(\epsilon L^{1/\nu})\label{eq:fu}.
\eea
where $f_q$, $f_\chi$ and $f_U$  are scaling functions, and $\nu$, $\beta$, $\gamma$ and $\alpha$ are the usual  critical exponents. 

From the Landau theory presented in Sec.~\ref{sec:landau}, we expect that the transition belongs to the universality class of 
three dimensional $O(3)$ model with cubic anisotropy. In $O(N)$ models with cubic anisotropy,
the phase transition is in the symmetric $O(N)$ universality class if $N<N_c$ and is in the cubic anisotropic universality class 
if $N>N_c$. Early work, using perturbative renormalisation group theory~\cite{1973-kw-jpa-modified,1973-w-jpc-critical,1973-a-prb-critical,1974-nkf-prl-renormalisation,1974-blz-prb-discussion}, high temperature series 
expansion~\cite{1981-fvc-prb-effect} and non-perturbative RG calculations~\cite{2002-tmvd-prb-randomly} suggests that $3<N_c<4$. 
Further work using RG calculations upto three loops~\cite{1988-ms-fls-cubic,1989-s-pla-critical}, four loops~\cite{1989-mss-ferro-critical,2000-v-prb-stability}, 
five loops~\cite{2000-ps-prb-five,1997-kts-prb-stability,1995-ks-plb-exact,1995-kt-prd-large} 
find $N_c\lesssim 3$, while six loop RG calculations suggest $N_c \approx 2.89$. Monte Carlo simulations are consistent with $N_c=3$~\cite{1998-ch-jpa-stability}. 
However, in three dimensions, the exponents for the model with cubic anisotropic critical are very close to the exponents for the Heisenberg 
model~\cite{2000-cpv-prb-component}. Therefore, we  use the exponents for the three-dimensional Heisenberg model to analyse the data. 
In the following, we check that the data near the critical point are consistent with these exponents.

\begin{figure}
\includegraphics[width=\columnwidth]{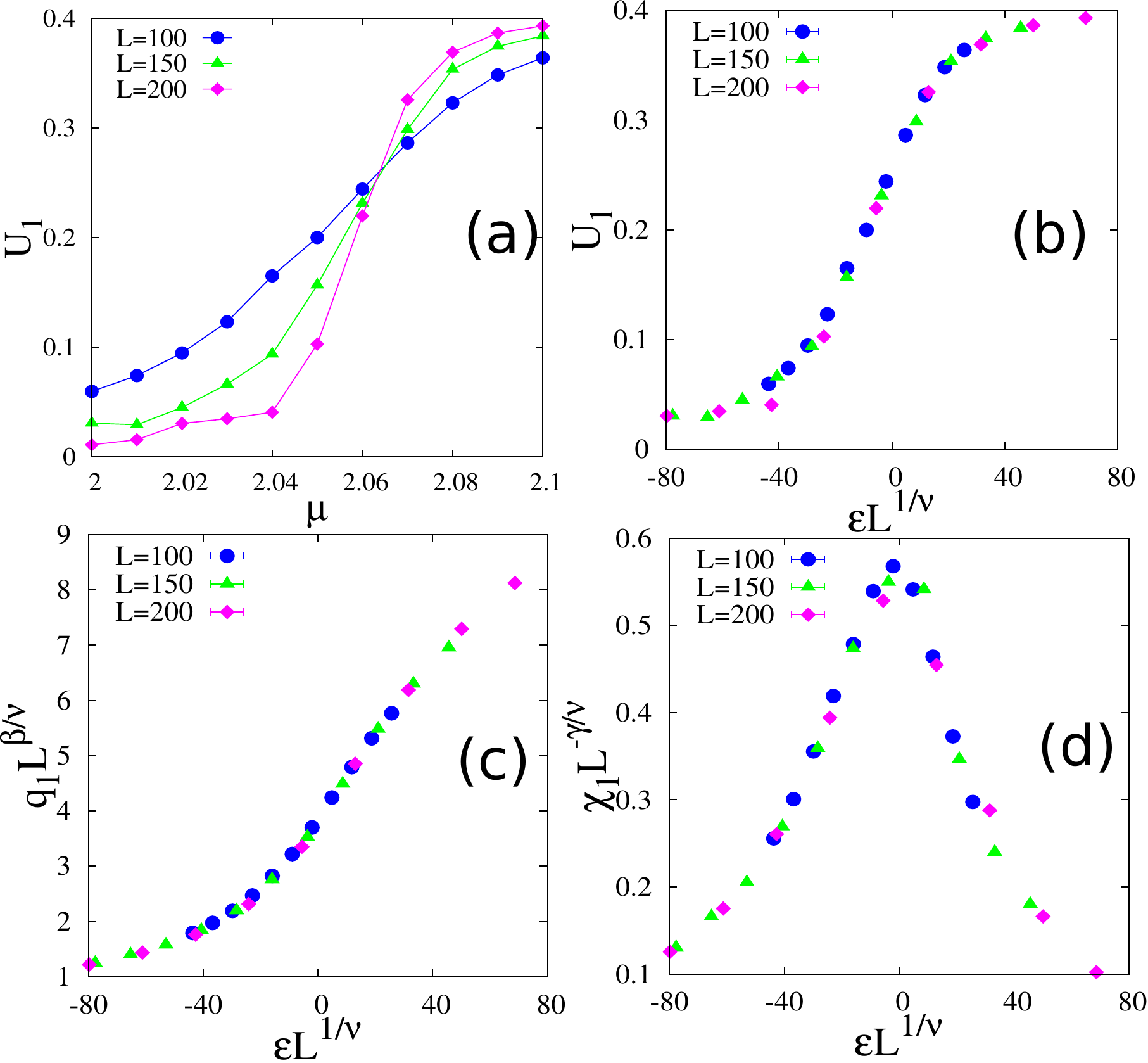}
\caption{\label{critical_heisenberg}(a) The variation of Binder cumulant $U_1$ with chemical potential $\mu$ for different system sizes. The data for  (b) Binder cumulant $U_1$, (c) order parameter $q_1$ and (d) $\chi_1$ for different system sizes collapse onto a single curve when scaled as in Eqs.~(\ref{eq:fq})-(\ref{eq:fu}) with the critical exponents of the three dimensional Heisenberg model: $\nu=0.704$, $\beta = 0.362$, and $\gamma=1.389$. }
\end{figure}

The critical point may be determined by the crossing point of the data for Binder cumulant for different system sizes. From this criterion, we
find that the critical parameters are $\mu \approx 2.063$ corresponding to $\rho \approx 0.718$ [see Fig.~\ref{critical_heisenberg}(a)]. 
The data for Binder cumulant, susceptibility, and order parameter for different system sizes collapse onto one curve when scaled as in 
Eqs.~(\ref{eq:fq})-(\ref{eq:fu}) with the critical exponents for Heisenberg model in three dimensions, as shown in Fig.~\ref{critical_heisenberg}. We use the numerical estimates for the critical exponents $\nu=0.704$, $\beta = 0.362$ and $\gamma=1.389$~\cite{1993-prb-hj-critical}. We conclude that the disordered-layered transition belongs to the universality class of the $O(3)$ model.

\section{\label{sec:transitions_2}Layered-Sublattice Transition}

In this section, we study the the nature of the layered-sublattice phase transition. We analyze the transition using the order parameter $q_3$, defined in Eq.~(\ref{eq:q3}), which is zero in the layered phase and non-zero in the sublattice phase. 

\begin{figure}
\includegraphics[width=\columnwidth]{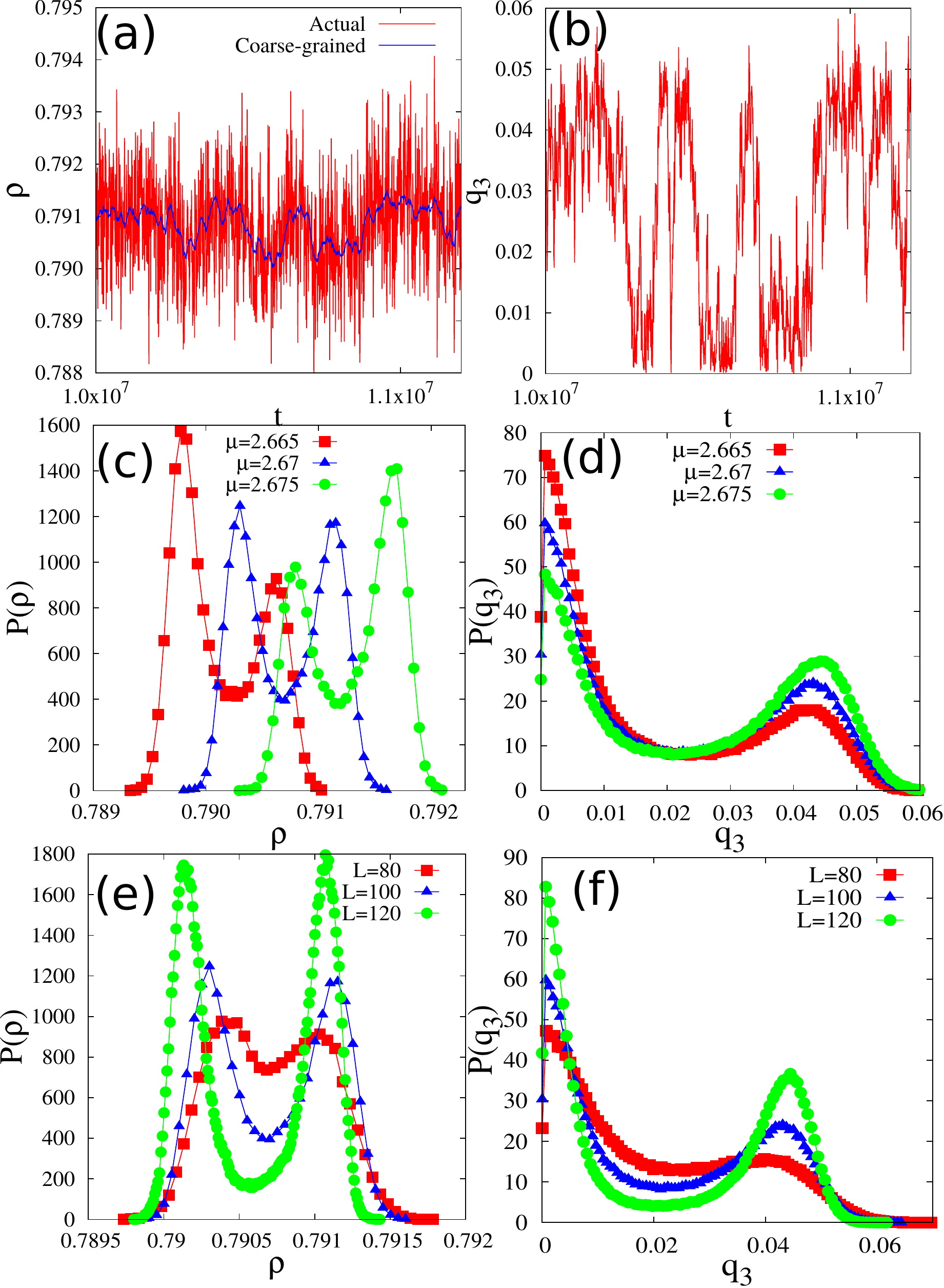}
\caption{\label{critical_2}  The time evolution of (a) density $\rho$ and (b) $q_3$ when $\mu=2.67$ and $L=100$. We have superimposed a running average of density, where each point has been averaged over 40 consecutive data points. The probability density function for (c) $\rho$ and (d) $q_3$ for different values of $\mu$ near the transition point for a system of size $L=100$. The probability density function for (e) $\rho$ and (f) $q_3$ for different values of $L$ at the transition point [$\mu=2.670$ for $L=80$, and $L=100$,  and $\mu=2.669$ for $L=120$].}
\end{figure}

Figures~\ref{critical_2}(a) and (b), show the time evolution of density and $q_3$ after equilibration. For clarity, we have also superimposed a running average of density, where each point has been averaged over 40 consecutive data points.  Both $\rho$ and $q_3$ exhibit two states, one in which density is higher and $q_3$ is non-zero and another where density is lower and $q_3$ is approximately zero. The system fluctuates in time between these two states. This is characteristic of a first order transition where both the sublattice and layered phases have the same free energy at the transition point. 

The probability distribution for density, $P(\rho)$, and $q_3$, $P(q_3)$ for three different values of $\mu$, close to the transition point, are shown in Figs.~\ref{critical_2}(c) and (d) respectively. Note that we have used the smoothened density to obtain the distribution. As the transition point is crossed, it can be seen that the distribution changes from having more area at the lower density to having more area  at the higher density. From the value of $\mu$ for which $P(\rho)$  has roughly same height, we conclude that  the critical chemical potential is $\mu\approx 2.67$.  Similar features may be seen for $P(q_3)$. Finally, the effect of system size on the distributions at the critical activity are studied in Figs.~\ref{critical_2}(e) and (f). It can be seen that the peaks become higher and sharper with increasing system size. The jump in density at the transition is $\approx 0.001$. These are again characteristic of a first order transition, and we conclude that the layered-sublattice transition is discontinuous. 

\section{\label{sec:transitions_3}Sublattice-Columnar Transition}

In this section, we study the the nature of the layered-sublattice phase transition. We analyze the transition using the order parameter $q_3$, as defined in Eq.~(\ref{eq:q3}). $q_3$ is zero in the columnar phase and non-zero in the sublattice phase. 

\begin{figure}
\includegraphics[width=\columnwidth]{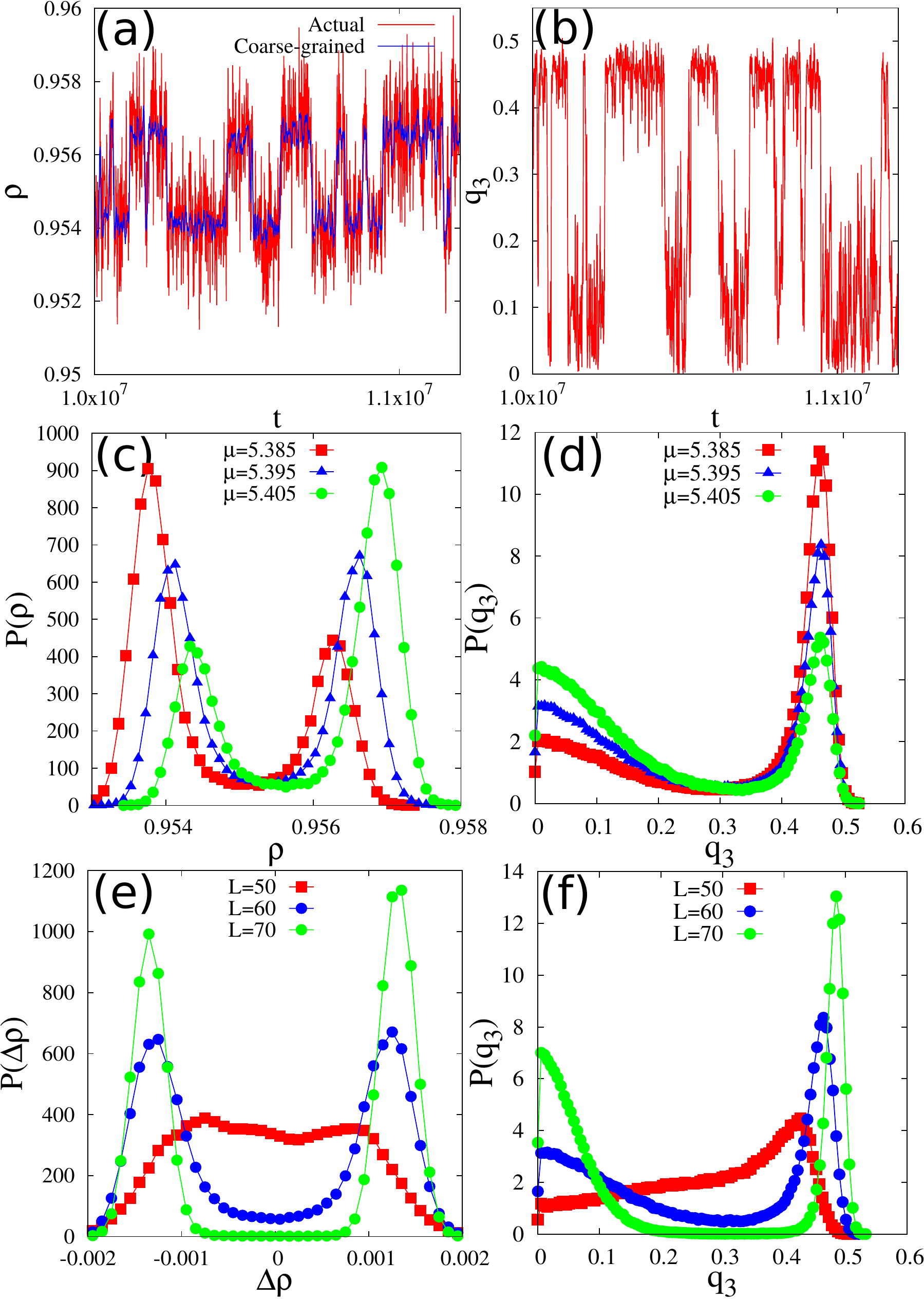}
\caption{\label{critical_3}  The time evolution of (a) density $\rho$ and (b) order parameter $q_3$ when $\mu=5.395$ and $L=60$. We have superimposed a running average of density, where each point has been averaged over 10 consecutive data points. The probability density function for (c) $\rho$ and (d) $q_3$ for different values of $\mu$ near the transition point for a system of size $L=60$. The probability density function for (e) $\Delta \rho =  \rho-\rho_c(L)$, where  $\rho_c(L)$ is midpoint between the two peaks in the distribution,   and (f) $q_3$,  for different values of $L$ at the transition point [$\mu_c=5.26, \rho_c \approx 0.9522$ for $L=50$,  $\mu_c=5.395, \rho_c \approx 0.9553$ for $L=60$ and $\mu_c=5.48, \rho_c\approx 0.9572$ for $L=70$].}
\end{figure}

Figures~\ref{critical_3}(a) and (b), show the time evolution of density and $q_3$ after equilibration. For clarity, we have also superimposed a running average of density, where each point has been averaged over 10 consecutive data points.  Both $\rho$ and $q_3$ exhibit two states, one in which density is higher and $q_3$ is non-zero and another where density is lower and $q_3$ is approximately zero. The system fluctuates in time between these two states. This is characteristic of a first order transition or co-existence,  where both the sublattice and columnar phases have the same free energy at the transition point. The jump in density across the transition is $\approx 0.0025$.

The probability distribution for density, $P(\rho)$, and $q_3$, $P(q_3)$ for three different values of $\mu$, close to the transition point, are shown in Figs.~\ref{critical_3}(c) and (d) respectively. Note that we have used the smoothened density to obtain the distribution. As the transition point is crossed, it can be seen that the distribution changes from having more area at the lower density to having more area  at the higher density. From the value of $\mu$ for which $P(\rho)$  has roughly same height, we conclude that  the critical chemical potential is $\mu\approx 5.395$ for $L=60$.  Similar features may be seen for $P(q_3)$. Finally, the effect of system size on the distributions at the critical activity are studied in Figs.~\ref{critical_3}(e) and (f). We find that the critical point has a strong  finite size dependence.  For instance, if the critical density $\rho_c(L)$ is defined as the midpoint between the two peaks in the distribution $P(\rho)$, then we find $\rho_c(50) \approx 0.9522$, $\rho_c(60) \approx 0.9553$, and $\rho_c(70) \approx 0.9572$. Since the difference in $\rho_c(L)$ are of the order of the jump in $\rho$ (approximately $0.0025$), we plot the the probability distribution of $\Delta \rho$ where $\Delta \rho = \rho-\rho_c(L)$. From Fig.~\ref{critical_3}(e), we find that, with increasing system size, the peaks become higher and sharper. The same features are seen for $P(q_3)$ [see Fig.~\ref{critical_3}(f)].  These are  characteristics of a first order transition, and we conclude that the sublattice-columnar transition is weakly first order.

\section{\label{sec:stability}Stability of Columnar Phase}

\begin{figure}
\centering
\includegraphics[width=\columnwidth]{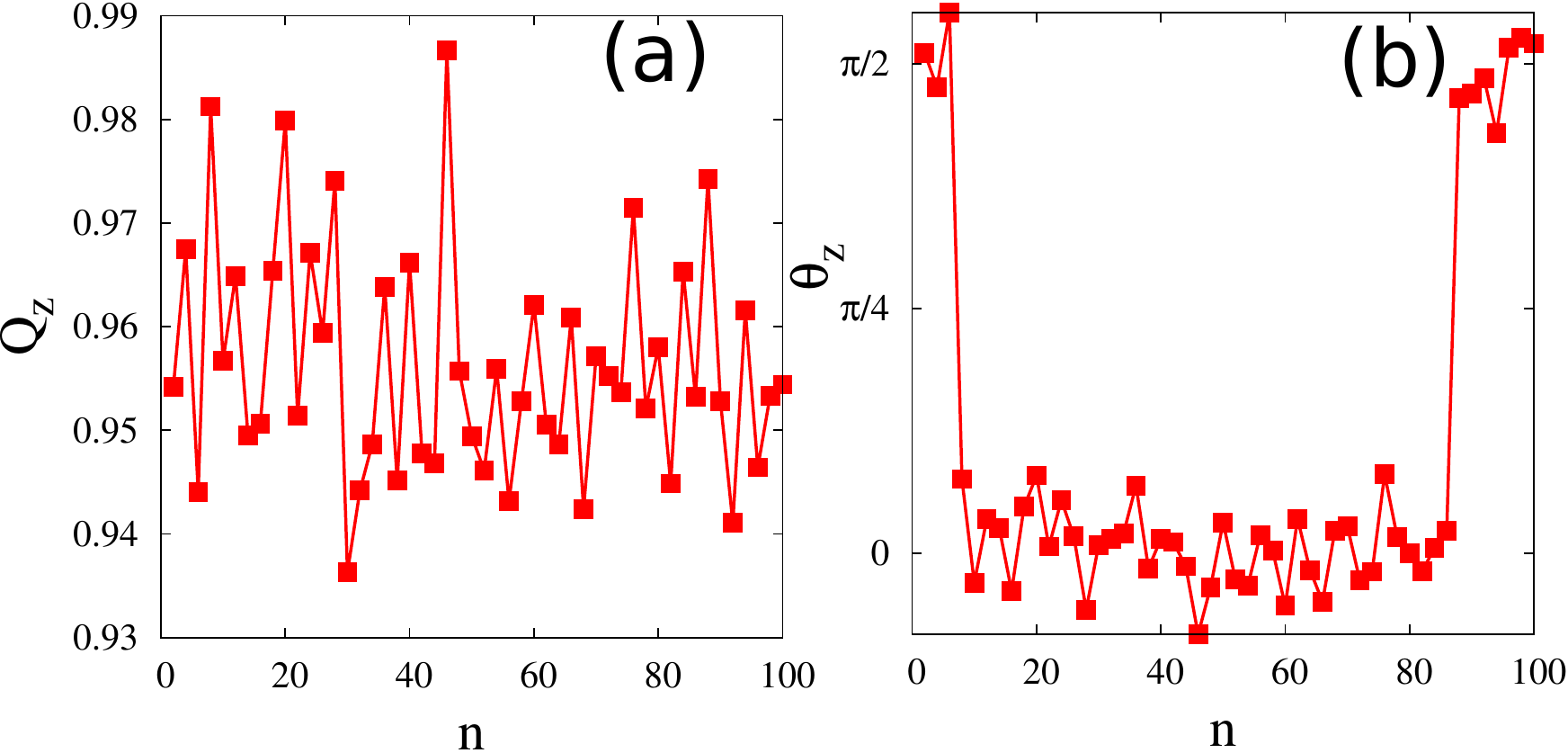}
\caption{\label{stuck_lay} Variation of (a) $Q_z$ and (b) $\theta_z$ [see Eq.~(\ref{eq:plane_lay}) for definition] with even $z$-planes. In the odd planes, there are very few cubes. The data are for $L=100$ and $\mu=6.0$ at $t=10^7$ Monte Carlo steps.}
\end{figure}

In our Monte Carlo simulations, we are unable to equilibrate the system efficiently in the high density phase for densities larger than $\rho \approx0.96$ for system sizes larger than $L\geq100$.  For these densities, we find that the system often gets stuck in very long-lived metastable states, which consist of layers of size $2\times L \times L$, each layer having a two dimensional columnar order. However, columnar order in consecutive layers may have different orientations. We illustrate with a typical  example that was obtained in simulations. Consider a system that is layered in the $z$-direction. We define the columnar order parameter for layer $n$, $\mathbf{Q}_{z}(n) = Q_{z}(n) \mathrm{e}^{i\theta_z(n)}$, as 
\be
Q_{z}(n)\,\mathrm{e}^{i\theta_z(n)}=
 \phi^z_{er}(n) - \phi^z_{or}(n)+ i\left[\phi^z_{ec}(n) - \phi^z_{oc}(n)\right],
\label{eq:plane_lay}
\ee
where $\phi^z_{er}(n)$ and $\phi^z_{or}(n)$ are the packing fraction of the cubes with heads lying on even and odd rows of the $n$-th plane respectively, while $\phi^z_{ec}(n)$ and $\phi^z_{oc}(n)$ are the corresponding  packing fractions on even and odd columns. For a layer with perfect columnar order, 
$\theta_z$ takes one of four values  $0$, $\pi/2$, $\pi$, $3 \pi/2$. Similar definitions hold for $Q_x(n)$ and $Q_y(n)$. In Fig.~\ref{stuck_lay}, we show the variation of $Q_{z}(n)$ and $\theta_z(n)$ of a configuration  that is layered in the $z$-direction (even planes are occupied), obtained  after equilibrating for $10^7$ Monte Carlo steps. It can be see that while the magnitude  remains constant across the even ayers, $\theta_z(n)=0$  or $\theta_z(n)=\pi/2$, showing that the columnar order in different planes have different orientations. We find that  the system remains stuck in this meta stable phase for upto and beyond $10^7$ Monte Carlo steps.

Though such metastable states exist, we will show below that the true equilibrium state is one with the same columnar order in all the layers. The presence of cubes that common to adjacent layers  tend to create an aligning interaction. This may be demonstrated though a perturbative calculation. This calculation is similar to the high density expansion developed for hard squares and rectangles~\cite{1967-bn-jcp-phase,2012-rd-pre-high,2015-nkr-jsp-high,2014-nr-pre-multiple,2016-ndr-epl-stability}.  

We start by assuming that the system is layered in the $z$-direction (even planes), and that $z$ is large enough so that there is perfect columnar order in each layer. We set up a perturbation expansion based on number of cubes with heads in odd planes. To do so, we introduce two activities, $z$ for cubes with heads on even planes and $z'$ for cubes with heads on odd planes. When $z'=0$, the layers are independent, and the problem reduces to that of $2\times2$ hard square lattice gas model. We will determine, to first order in perturbation theory, the difference in free energy between a state  in which all planes have even-row order and states in which one plane is misaligned with either odd-row order or column-ordered.  We will denote these states by $S_\alpha$, whose partition function $\lm^{\alpha}(z,z')$ and the free-energy  $F^{\alpha}(z,z')=-\ln \lm^{\alpha}(z,z') $  can be formally written as
\bea
\lm^{\alpha}(z,z') &=& \lm^{\alpha}_0(z) + z'\,\lm^{\alpha}_1(z) + \mathcal{O}(z'^2),\\
F^{\alpha}(z,z') &=& F^{\alpha}_0(z)  - z'\,\frac{\lm^{\alpha}_1(z)}{\lm^{\alpha}_0(z)}+ \mathcal{O}(z'^2).
\eea

For $S_{||}$, for which all planes have even-row order, the  partition function, when there are no defect cubes (cubes with heads on odd planes), is
\be
\lm^{||}_0(z) = [\pbc^{L/2}]^{L/2},
\ee
where $\Omega_p(L)$ is the partition function of a periodic column  of size $2\times2\times L$. Consider a single defect cube that is placed in any of the 
$L/2$ odd planes. Wiithin a plane, it can choose any one of  $L^2$ sites. The partition function for  $S_{||}$ in the presence of one defect cube is
\be
\lm^{||}_1(z)=\frac{L^3}{4} [\pbc]^{\frac{L^2}{4}}\left( \left[ \frac{\obc{L\!-\!2}}{\pbc}\right] ^2\!\!+\left[\frac{\obc{L\!-\!2}}{\pbc}\right]^4\right),
\label{1st_corr}
\ee
where $\Omega_o(L-2)$ is the partition function of an open column of size $2\times2\times (L-2)$. In Eq.~(\ref{1st_corr}), the first term represents the correction coming when the head of the defect cube is placed on an even row and second term corresponds to when the head of the defect cube is placed on an odd row.

Consider now a  state $S_{ro}$ in which one of the planes (say $z=0$)  is odd-row ordered. Since the partition function with no defect is identical to that for $S_{||}$,   the difference in partition function appears only in the first-order correction term: 
\begin{align}
& \lm^{ro}_1(z) = [\pbc^{L/2}]^{\frac{L}{2}-2} \bigg[\left(\frac{L}{2}-2\right) \frac{L^2}{2}\times\nonumber\\
& \left\{ [\obc{L-2}\pbc^{\frac{L}{2}-1}]^2 + [\obc{L-2}^2\pbc^{\frac{L}{2}-2}]^2\right\} \nonumber\\
& + 2L^2 \obc{L-2}^3\pbc^{L-3}\bigg].
\end{align}
The difference in free-energies, $\Delta F^{||, ro}(z, z')$, may be written as
\bea
\Delta F^{||, ro}(z,z') &=& F^{ro}(z,z') - F^{||}(z,z'),\nonumber\\
&=& z'\left(\frac{\lm^{||}_1(z) - \lm^{ro}_1(z)}{\lm^{||}_0(z)}\right). 
\label{free1}
\eea
Simplifying Eq.~(\ref{free1}), we obtain
\be
\Delta F^{||, ro}(z,z') =L^2z'\left(\left[\frac{\obc{L-2}}{\pbc}\right]^2 - \left[\frac{\obc{L-2}}{\pbc}\right]\right)^2.
\label{delta1}
\ee
Since the right hand side is a perfect square, $\Delta F^{||, ro}(z,z') > 0$ for any $L$. Thus, the state with one misaligned row-ordered state has higher free energy, and we conclude that introduction of defect cubes results in an effective aligning interaction that tends to make all the planes have columnar order with the same orientation.

The large $z$ behaviour of $\Delta F^{||, ro}(z,z')$ may be determined  by noting that for large $L$, $\pbc = a_p\lambda^L$ and $\obc{L-2} = a_o\lambda^{L-2}$  where $\lambda$ is the largest root of the equation $x^2-x-z=0$. From Ref.~\cite{2016-ndr-epl-stability}, we obtain $a_p=1$, $a_o = \lambda/(2\lambda-1)$ and $\lambda = (1+\sqrt{1+4z})/2$. Setting $z'=z$ and evaluating in the limit of $z\gg 1$, we obtain
\be
\Delta F^{||, ro}(z) = \frac{L^2}{4z}  + \mathcal{O}(z^{-3/2}).\label{eq:deltaf}
\ee 

It is straightforward to generalize this calculation to the state $S_{ce}$  in which one of the planes has column-order. We omit the calculation, but we obtain a similar increase in free energy when a layer is misaligned.

We now ask whether metastable states, as seen in Fig.~\ref{stuck_lay}, are due to finite size effects or due to the algorithm being unable to equilibrate the system at high densities within available computer time.  Though a misaligned plane results in a rise in free energy as Eq.~(\ref{eq:deltaf}), there is a gain in entropy $\ln 2$ per column when there are no defect cubes. Thus, we can identify a crossover length $L^*$ at which the free energy gained by alignment of a plane is balanced by the entropy lost due to alignment of such a plane. Equating the two free energies, $\Delta F^{||, ro}(z,z')\sim \ln 2$, we obtain
\be
L^* = \sqrt{4 z \ln 2}.
\ee
For the metastable state shown in Fig.~\ref{stuck_lay},  $z= 403.43$ for the state , we obtain $L^* \approx 33$. For system sizes smaller than this length, the misaligned phases are favoured, but are a finite size effect. However, since the system lengths that we have simulated are much larger than $L^*$, we conclude that the presence of such metastable states are due to an inability of the Monte Carlo algorithm to equilibrate states with misaligned planes for large $z$, due to large entropic barriers.

\section{\label{sec:discussion}Summary and conclusions}

\begin{figure}
\includegraphics[width=\columnwidth]{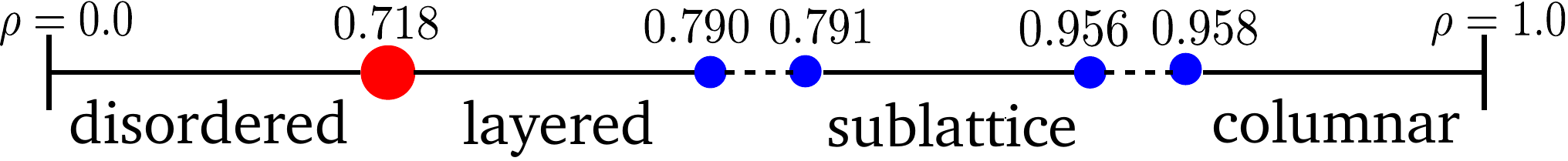}
\caption{\label{schematic} Numerically obtained phase diagram for  $2\times2\times2$ hard cubes. The red dot represents a continuous transition and the dotted lines represent regions of coexistence.} 
\end{figure}

In this paper we studied the phases and the phase transitions in a system of $2\times2\times2$ hard cubes on a three dimensional cubic lattice. We show the existence of four different phases. In order of increasing density, these are a disordered phase, a layered phase in which the system breaks up into $L/2$ interacting slabs of size $2 \times L\times L$ each having fluid-like order, a solid-like sublattice phase, where the cubes preferentially occupy one sublattice, and a columnar phase in which the system breaks up into $L^2/4$ columns of dimension $2 \times 2 \times L$ with a fluid-like order within a column.  The disordered-layered transition is shown to be a continuous transition that is consistent with the universality class of the three dimensional $O(3)$ model with cubic anisotropy. The other two transitions -- layered-sublattice and sublattice-columnar -- are shown to be discontinuous. The phase diagram is summarized in Fig.~\ref{schematic}.

We formulated a Landau theory, consistent with the symmetries of the system, that is able to describe all the phases seen in the Monte Carlo simulations.
Within the minimal functional, as described in Eq.~(\ref{eq:landau}), based only on the layering vector $\mathbf{L}$, we find that the columnar phase is unstable for the full range of the parameters, while it predicts the existence and stability of  disordered, layered and sublattice phases. It also predicts the disordered-layered and disordered-sublattice transitions to be continuous and belong to the universality class of  the three-dimensional $O(3)$ model with cubic anisotropy.  The layered-sublattice transition is predicted to be first order. To obtain a stable columnar phase, we extended the free energy functional to explicitly depend on  the columnar vector $\mathbf{C}$ [see Eq.~(\ref{eq:landau_coupled})]. Within  this extended functional, it is possible to show the existence and stability of all the different phases observed in simulations.

The results in this paper are not consistent with the theoretical predictions of density functional theory.
Density functional theory predicts that $2\times2\times2$ hard cubes cubes undergo  transitions from a disordered phase to layered phase to a columnar phase at high densities~\cite{2003-jcp-lc-phase}. However, it does not predict the sublattice phase, that is seen in our Monte Carlo simulations. Understanding why the theory fails, and how it should be modified to give the correct predictions is a promising area for future study. For $6\times 6 \times 6$ cubes, the theory predicts a transition from a disordered to solid to two types of columnar phase~\cite{2003-jcp-lc-phase}. Testing these predictions in simulations  would also be of interest.

The results in this paper are also in contradiction to earlier Monte Carlo simulations~\cite{2005-p-jcp-thermodynamic}, wherein no phase transitions were found even though the simulations were performed close to full packing (in Ref.~\cite{2005-p-jcp-thermodynamic}, the problem of cubes correspond to 
$\sigma=2$). This discrepancy could be due to small system sizes that were studied ($L=18,24$) in Ref.~\cite{2005-p-jcp-thermodynamic} compared to the systems sizes studied in the current paper ($L$ upto $200$).

The existence of a sublattice phase is quite surprising. It would appear that as you introduce vacancies at full packing, the sublattice phase gets destabilised in favour of the columnar phase. However, a larger number of vacancies somehow stabilises the sublattice phase. Also, it implies that the interactions between the different layers in the layered phase are not weak. If they were weak, then we would expect that once the system becomes layered, the problem becomes effectively a problem of hard squares in two dimensions. This lower dimensional system does not exhibit a sublattice phase.

In the continuum, the system of parallel hard cubes undergoes a continuous freezing transition from a disordered fluid phase to a solid phase as density is increased~\cite{1998-pre-j-melting,2001-jcp-gm-closer}, consistent with theoretical predictions using density functional theory~\cite{2012-jcp-bdr-free}. The continuum limit may be reached by determining the phase diagram for $k\times k \times k$ cubes and extrapolating for large $k$. Preliminary simulations for $3\times3\times3$ cubes suggest that the sublattice phase does not exist, but the layered and columnar phases exist. Thus, for larger $k$, the layered-columnar transition would appear to be similar to the disordered-columnar transition in $k\times k$ hard squares. For this model, high density expansions suggest that the critical density tends to an asymptotic value that is less than one for large $k$~\cite{2015-nkr-jsp-high}. If this were true, the continuum problem should have two transitions. Re-examining the problem of parallel hard cubes in the continuum  is a promising area for future study. 

\begin{acknowledgments}
The simulations were carried out in single node cluster machine Nandadevi (2 x Intel Xeon E5-2667  3.3 GHz) using OpenMP parallelisation. DD's research was partially supported by the J. C. Fellowship, awarded by the Department of Science and 
Technology, India, under the grant DST-SR-S2/JCB-24/2005.
\end{acknowledgments}


%

\end{document}